\documentclass{aa}
\usepackage{color}
\usepackage{colortbl}
\usepackage{multirow}
\usepackage{dcolumn}
\usepackage{amsmath}
\usepackage{amssymb}
\usepackage{graphicx}
\usepackage{epsfig}
\usepackage{changebar}
\usepackage{natbib}
\usepackage{txfonts}
\usepackage{relsize}

\usepackage[figuresright]{rotating}

\newcolumntype{d}[1]{D{.}{\cdot}{#1}}

\newcolumntype{.}{D{.}{.}{-1}}

\newcommand{\lsun}{L$_\odot$}

\newcommand{\mum}{$\mu$m}

\begin{document}
   \title{The RMS Survey}

  \subtitle{6~cm continuum VLA observations towards candidate massive YSOs in the northern hemisphere\thanks{Tables 3, 4 and 5 and full versions of Figs. 2, 4 and 7 are only available in electronic form at the CDS via anonymous ftp to cdsarc.u-strasbg.fr (130.79.125.5) or via http://cdsweb.u-strasbg.fr/cgi-bin/qcat?J/A+A/.}}

   \author{J.~S.~Urquhart
          \inst{1,2}  
			 \and
			 M.~G.~Hoare
			 \inst{1}
			 \and
			 C.~R.~Purcell
			 \inst{3}
			 \and
			 S.~L.~Lumsden
			 \inst{1}
			 \and
			 R.~D.~Oudmaijer
			 \inst{1}
			 \and
			 T.~J.~T.~Moore
			 \inst{4}
			 \and
			 A.~L.~Busfield
			 \inst{1} 
			 \and
			 J.~C.~Mottram
			 \inst{1,5} 
			 \and
			 B.~Davies
			 \inst{1} 
          }

     \offprints{J. S. Urquhart: jsu@ast.leeds.ac.uk}

   \institute{School of Physics and Astrophysics, University of Leeds, Leeds, LS2~9JT, UK 
         \and Australia Telescope National Facility, CSIRO, Sydney, NSW 2052, Australia   
         \and 
			Jodrell Bank Centre for Astrophysics, University of Manchester, Oxford 
Road, Manchester, M13 9PL, UK
         \and    Astrophysics Research Institute, Liverpool John Moores University, Twelve Quays House, Egerton Wharf, Birkenhead, CH41~1LD, UK
          \and School of Physics, University of Exeter, Exeter, EX4 4QL, UK   
             }

   \date{}

\abstract
   {The Red MSX Source (RMS) survey is an ongoing multi-wavelength observational programme designed to return a large, well-selected sample of massive young stellar objects (MYSOs). We have identified $\sim$2000 MYSO candidates located throughout the Galaxy by comparing the colours of MSX and 2MASS point sources to those of known MYSOs. The aim of these follow-up observations is to identify other objects with similar colours such as ultra compact (UC) HII regions, evolved stars and planetary nebulae (PNe) and distinguish between genuine MYSOs and nearby low-mass YSOs. }      
   {To identify the populations of UCHII regions and PNe within the sample and examine their Galactic distribution.} 
   {We have conducted high resolution radio continuum observations at 6~cm towards 659 MYSO candidates in the northern hemisphere ($10\degr< l < 250\degr$) using the Very Large Array (VLA). These observations have a spatial resolution of $\sim$1--2\arcsec\ and typical image r.m.s. noise values of $\sim$0.22~mJy -- sensitive enough to detect a HII region powered by B0.5 star at the far side of the Galaxy. In addition to these targeted observations we present archival data towards a further 315 RMS sources extracted from a previous VLA survey of the inner Galaxy.}
   {We present the results of radio continuum observations made towards 974 MYSO candidates, 272 ($\sim$27\% of the observed sample) of which are found to be associated with radio emission above a 4$\sigma$ detection limit ($\sim$1~mJy). Using results from other parts of our multi-wavelength survey we separate these RMS-radio associations into two distinct types of objects, classifying 51 as PNe and a further 208 as either compact or UC HII regions. Including all HII regions and PNe identified either from the literature or from the multi-wavelength data these numbers increase to  391 and 79, respectively. Using this well selected sample of HII regions we estimate their Galactic scale height to be 0.6\degr. In addition to the RMS-radio associations we are able to set upper limits on the radio emission of $\leq$1~mJy for the 702 non-detections, which is below the level expected if they had already begun to ionise their surroundings. }
  {Using radio continuum and archival data we have identified 79 PNe and 391 HII regions within the northern RMS catalogue. We estimate the total fraction of contamination by PNe in the RMS sample is of order 10\%. The sample of HII regions is probably the best representation to date of the Galactic population of HII regions as a whole.}
\authorrunning{J. S. Urquhart et al.}
\titlerunning{Radio observations of MYSO candidates}
\maketitle

\section{Introduction}

Massive young stellar objects (hereafter MYSOs) are one of the earliest phases in the life of OB stars when fusion has most likely started in the core, but they have not yet begun to ionise their surroundings to form an HII
region. They possess strong ionised stellar winds and drive powerful bipolar
molecular outflows. This brief ($\sim$$10^{4-5}$ years) phase is clearly
crucial to our understanding of how these massive young stars form
since during this time any ongoing accretion will be disrupted and the
final mass of the star may be set. In addition, the winds, outflows and
eventual HII regions have an important feedback role in the fate of
the rest of the molecular cloud and subsequent generations of high-mass star formation (e.g., \citealt{moore2007,urquhart2007d,deharveng2009}). 

The main difficulty in studying MYSOs
is their relative rarity.  The $\sim$30 well-known MYSOs form a
heterogeneous sample (\citealt{henning1984}), often discovered
serendipitously, and are still largely being found this way
(e.g., \citealt{shepherd1998, cesaroni1999}).  Unfortunately the known
sample is too small to test many aspects of massive star formation
theories and may be unrepresentative of the class as a whole.
The obvious solution is to carry out a systematic and statistically complete
search for MYSOs.

MYSOs are brightest in the mid- to far-infrared where most of the
bolometric luminosity emerges after reprocessing by the surrounding
dust cloud. Hence, any statistically complete sample must
therefore be initially selected from such data.  Searches for MYSOs using mid-
to far-infrared data have been made using the IRAS all-sky survey
(\citealt{campbell1989, chan1996,molinari1996, sridharan2002}), but the concentrations of massive stars within $\sim$$\pm$2\degr\ of the Galactic plane means there is considerable source confusion due to the large IRAS beam
($\sim$2--5\arcmin\ at 100 $\mu$m). The Midcourse Space Experiment
(MSX) satellite recently completed a much higher spatial resolution
survey ($\sim$18\arcsec) of the Galactic plane ($|b|<5^\circ$) at 8,
12, 14 and 20~$\mu$m (\citealt{price2001}), which completely
supersedes IRAS mid-infrared data. We have developed 
colour-selection criteria by comparing the colours of sources from the
MSX and 2MASS point source catalogues to those of known MYSOs. This
has delivered a sample of approximately 2000 candidate MYSOs spread
throughout the Galaxy ($|b|<5$\degr; \citealt{lumsden2002}).

MYSOs are very red, but unfortunately several other kinds of objects
have very similar infrared colours, especially compact HII regions and
some planetary nebulae (PNe), post-AGB stars, dusty red supergiants
as well as nearby low- and intermediate-mass YSOs. The Red MSX Source
(RMS) survey is a multi-wavelength programme of follow-up
observations designed to distinguish between genuine MYSOs and these
other kinds of embedded or dusty objects
(\citealt{hoare2005,mottram2006,urquhart2007c}), and to compile a
database of complementary multi-wavelength data with which to study
their properties.\footnote{http://www.ast.leeds.ac.uk/RMS} These
include: molecular line observations to obtain kinematic distances
which are crucial for calculating luminosity, and hence distinguishing
between nearby low- and intermediate-mass YSOs and genuine MYSOs
(\citealt{urquhart_13co_south,urquhart_13co_north}); mid-infrared
imaging to identify genuine point sources, obtain accurate astrometry,
and avoid excluding MYSOs located near compact and ultra-compact (UC)
HII regions (e.g., \citealt{mottram2007}); and near-infrared
spectroscopy (e.g., \citealt{clarke2006}) to distinguish between MYSOs
and evolved stars.

Another aspect of the RMS follow-up campaign is to obtain high resolution
radio continuum observations to identify compact HII regions and PNe
that contaminate the sample. Nearby ($\sim$1~kpc) MYSOs have an
integrated radio flux from their ionised stellar winds of a few mJy at
most (\citealt{hoare1994, rodriguez1994, hoare1996, hoare2002}),
whilst most compact HII regions have much greater radio fluxes
(\citealt{kurtz1994,giveon2005a}). High spatial resolution is needed
since very often MYSOs are located close (few arcseconds) to compact
HII regions and we must be sure not to exclude these since that would
bias against sites of sequential star formation.  Although we
want to eliminate compact HII regions as confusing sources, the RMS survey
will also deliver a large and well-selected sample of these too.

In a recent paper (\citealt{urquhart_radio_south}; hereafter Paper~I)
we presented the results of a set of high resolution radio
observations towards 826 RMS sources located in the southern
hemisphere (i.e., 230\degr$<l<$350\degr). These observations were made
at 4.8 and 8.6~GHz with the Australia Telescope Compact Array (ATCA)\footnote{The Australia Telescope Compact Array is
funded by the Commonwealth of Australia for operation as a National
Facility managed by CSIRO.}
and had a 4$\sigma$ sensitivity of $\sim$1~mJy and spatial resolution
of $\sim$2\arcsec. These observations resulted in radio emission being detected towards $\sim$200 RMS sources, a detection
rate of $\sim$25\%. The majority of these, due to their morphologies
and location close to the Galactic mid-plane, are thought to be
compact and UCHII regions. In this paper we present a set of
complementary high resolution observations made with the Very Large
Array (VLA) towards a further 659 RMS sources located in the northern
hemisphere (10\degr$<l<$250\degr) for which previous high resolution
radio observations were not available.

The structure of this paper is as follows: in Sect.~2 we outline our
source selection, observational and data reduction procedures, and present
a catalogue of the radio sources detected. In Sect.~3 we introduce additional
archival data and present the criterion used for associating radio emission
with RMS sources. We present the results and statistical analysis in
Sect.~4. In Sect.~5 we distinguish between HII regions and PNe and
discuss the Galactic distributions of these two populations. Finally
in Sect.~\ref{sect:summary} we present a summary of the results and
highlight our main findings.

\section{6~cm radio continuum observations}
\label{sect:observations}

\subsection{Source selection}
 
Our colour selection criteria identified $\sim$2000 MYSO candidates
spread throughout the Galaxy in a latitude range of
$|b|<$5\degr (\citealt{lumsden2002}). Sources toward the Galactic centre were excluded
($|l|<10\degr$) due to confusion and difficulties in calculating
kinematic distances.  Of these, 1101 are observable by the VLA.

In order to reduce the number of observations required, and to avoid sources that have been observed as part of other programmes,
we took the following steps: 1) conducted a literature search to
identify and remove well known sources whose identification was clearly established; 2) we cross-matched our source list with previous
high resolution surveys (i.e.,
\citealt{wood1989,kurtz1994,walsh1998,sridharan2002}) eliminating
sources for which data were already available; 3) conducted a
nearest-neighbour search to identify small groups of sources that were
close enough to be observed in a single field.

From a literature search we eliminated 127 sources including 74
evolved stars, of which 28 are identified as PNe, 18 well known
sources that have previously been studied in detail, and a further 35
YSO/HII regions that have been observed as part of the four radio
surveys mentioned in the previous paragraph. In addition to these
sources we also excluded 315 RMS sources that are located within the
region of the inner Galaxy surveyed by \citet{becker1994} in their 5~GHz VLA survey. We present a summary of this survey and identify the radio data associated with the RMS sources within this region in
Sect.~\ref{sect:magpis}. In the rest of this section we will describe
the observational set up, data reduction and source extraction
procedures for these targeted VLA observations.

\subsection{VLA observational set up}

\begin{table}
\begin{center}
\caption{Summary of observation dates, array configurations and numbers of sources observed.}
\label{tbl:observation_dates}
\begin{minipage}{\linewidth}
\begin{tabular}{ccccc}
\hline
\hline
VLA Project	& Date& Array	& Bandwidth& Num. \\
Id.	& (dd/mm/yyyy)& Config.& (MHz)& Obs. \\
\hline

AH786 	& 23/07/2002 & B & $2\times 25$ & 230 \\
 		& 25/07/2002 & B & $2\times 25$ & 125 \\
\hline 		
AH832 	& 22/09/2003 & BnA & $2\times 50$ & 72 \\
 		& 26/11/2003 & B & $2\times 50$ & 230 \\
 		& 13/12/2003 & B & $2\times 50$ & 210 \\
\hline
AH869 	& 13/02/2005 & BnA& $2\times 50$ &65 \\
 		& 03/05/2005 & B & $2\times 50$ & 122 \\
 		& 08/05/2005 & B & $2\times 50$ & 51 \\
\hline
\end{tabular}\\

\end{minipage}
\end{center}
\end{table}

Radio continuum observations were made during three seasons between
July 2002 and May 2005 using the Very Large Array (VLA;
http://www.vla.nrao.edu/) of the National Radio Astronomy
Observatory.\footnote{The National Radio Astronomy Observatory is a
facility of the National Science Foundation operated under
co-operative agreement by Associated Universities, Inc.}  The
observations were made at a frequency of $\sim$5~GHz, corresponding to a wavelength of $\sim$6~cm. Two array configurations were used; the B array for fields north of $-$10\degr\ declination (the majority of the observations), and where possible, the hybrid BnA array for sources south of ~$-$10\degr\ declination as it results in a more circular beam for low elevation sources. These configurations provide a spatial resolution of $\sim$1--2\arcsec.  To improve sensitivity and reduce bandwidth smearing two separate bandpasses were used (i.e., 2IFs) located on either side of the observing frequency. The observations made in 2002 used $2\times25$~MHz bandwidth with all subsequent observations using a $2\times50$~MHz bandwidth. In
Table~\ref{tbl:observation_dates} we present a summary of the
observation dates, VLA project codes, array configurations and numbers
of sources observed. 

Sources were grouped by position into small blocks of between 8--10
sources, with each source being observed for approximately 2--3
minutes within the block. To correct for fluctuations in the phase and amplitude of these data, caused by atmospheric and instrumental effects, each block was
sandwiched between two short observations of a nearby phase calibrator
(typically 2--3 minutes depending on the flux density of the
calibrator). The primary flux calibrator 1331+305 was observed once
during each set of observations to allow the absolute calibration of
the flux density. The observational parameters are summarised in
Table~\ref{tbl:radio_setup}.

\begin{table}
\begin{center}
\caption{Summary of the observational parameters.}
\label{tbl:radio_setup}
\begin{minipage}{\linewidth}
\begin{tabular}{lc}
\hline
\hline
Parameter&   \\
\hline
Frequency \dotfill &  4.86 GHz\\
Primary beam \dotfill & 4.4$^{\prime}$ \\
Synthesised beam\footnote{Declination and hour angle dependent.}\dotfill   & $\sim$1.5$^{\prime\prime}$ \\
Largest well imaged structure$^a$\dotfill & $\sim$20$^{\prime\prime}$ \\
Theoretical image r.m.s. \dotfill & $\sim$0.14--0.19 mJy beam$^{-1}$\\
Typical image rms\footnote{The stated r.m.s. values have been estimated from emission free regions close to the centre of imaged fields.}\dotfill &  0.2 mJy beam$^{-1}$\\
Image pixel size \dotfill &  0.3$^{\prime\prime}$ \\

\hline
\end{tabular}\\

\end{minipage}
\end{center}
\end{table}

In total 1105 observations were made towards 1068 fields, with a
couple of additional observations being made towards a small number of
the more complicated emission regions to improve \emph{uv} coverage
and reduce confusion. The field names and positions are presented in
Table~3.

\begin{table*}
\begin{center}
\caption{Field positions and restoring beam parameters.}
\label{tbl:fields_positons}
\begin{tabular}{lccccr}
\hline
\hline
Field Name	&	RA		&	Dec.		&	Field		&	Beam Size	  \\
			&	(J2000)	&	(J2000)	&	r.m.s. (mJy) &	Maj $\times$ Min (PA) (\arcsec,\arcsec,\degr) 	  \\
\hline
G010.0026$-$02	&	01:13:01.73	&	-21:16:14.0	&	0.14	&	2.0 $\times$ 1.1	($-$08.8)\\
G010.0712$-$00	&	01:12:37.93	&	-20:25:55.0	&	0.29	&	2.8 $\times$ 1.0	(+21.2)\\
G010.1434$-$00	&	01:12:38.54	&	-20:22:10.0	&	1.42	&	2.6 $\times$ 1.1	($-$22.8)\\
G010.1960$-$00	&	01:12:39.02	&	-20:19:30.0	&	0.34	&	3.1 $\times$ 0.9	(+21.7)\\
G010.3844+02	&	01:12:01.51	&	-18:52:08.0	&	0.11	&	1.4 $\times$ 0.7	($-$72.5)\\
G010.3920+00	&	01:12:26.28	&	-19:41:05.0	&	0.17	&	5.2 $\times$ 1.2	($-$43.2)\\
G010.4984$-$00	&	01:12:46.66	&	-20:13:35.0	&	0.12	&	1.3 $\times$ 0.8	($-$78.3)\\
G010.5067+02	&	01:12:02.30	&	-18:45:17.0	&	0.12	&	1.4 $\times$ 0.7	($-$72.7)\\
G010.5888$-$00	&	01:12:42.09	&	-19:58:32.0	&	1.34	&	0.7 $\times$ 0.7	($-$45.0)\\
G010.5960$-$00	&	01:12:49.00	&	-20:11:25.0	&	0.09	&	0.7 $\times$ 0.7	(+00.0)\\
G010.8407$-$03	&	01:13:35.33	&	-21:21:52.0	&	0.15	&	1.2 $\times$ 0.8	(+81.7)\\
G010.8837+01	&	01:12:16.16	&	-18:47:11.0	&	0.18	&	1.3 $\times$ 0.7	($-$84.2)\\
G010.9257+01	&	01:12:14.60	&	-18:41:09.0	&	0.17	&	1.3 $\times$ 0.7	($-$83.9)\\
G011.0054+00	&	01:12:25.68	&	-18:57:46.0	&	0.18	&	5.0 $\times$ 1.2	($-$43.2)\\
G011.0421$-$01	&	01:12:57.42	&	-19:57:07.0	&	0.14	&	2.0 $\times$ 1.1	($-$10.3)\\
G011.0909+00	&	01:12:33.90	&	-19:08:06.0	&	0.15	&	2.6 $\times$ 1.1	($-$22.6)\\
G011.1827+03	&	01:11:54.38	&	-17:42:30.0	&	0.20	&	1.3 $\times$ 0.7	($-$81.9)\\
G011.2230+00	&	01:12:34.82	&	-19:00:52.0	&	0.15	&	2.5 $\times$ 1.1	($-$21.9)\\
G011.2689$-$00	&	01:12:54.00	&	-19:35:03.0	&	0.17	&	4.7 $\times$ 1.2	($-$42.1)\\
G011.3161$-$01	&	01:13:02.28	&	-19:47:42.0	&	0.14	&	3.6 $\times$ 0.9	(+25.4)\\
G011.3252$-$01	&	01:13:08.86	&	-19:59:36.0	&	0.11	&	1.3 $\times$ 0.8	($-$88.4)\\
G011.3743$-$00	&	01:12:55.00	&	-19:29:46.0	&	0.12	&	1.2 $\times$ 0.8	($-$89.8)\\
G011.3757$-$01	&	01:13:07.36	&	-19:53:20.0	&	0.18	&	1.3 $\times$ 0.8	($-$86.2)\\
G011.4136+00	&	01:12:36.31	&	-18:50:43.0	&	0.11	&	2.2 $\times$ 1.0	(+17.6)\\
G011.4201$-$01	&	01:13:07.79	&	-19:51:07.0	&	0.11	&	1.3 $\times$ 0.8	($-$86.6)\\

\hline
\end{tabular}\\
\end{center}
Notes: Only a small portion of the data is provided here, the full table is available in electronic form at the CDS via anonymous ftp to cdsarc.u-strasbg.fr (130.79.125.5) or via http://cdsweb.u-strasbg.fr/cgi-bin/qcat?J/A+A/.

\end{table*}

\subsection{Data reduction}

The calibration and reduction of these data were performed using a
combination of tasks from the AIPS and Obit software packages. Raw
data were first inspected by eye and obvious phase and amplitude
excursions were flagged as bad. Data from each observing session
was calibrated using the standard AIPS tasks\footnote{See Section~4 of
the AIPS cookbook, available from http://www.nrao.edu/aips.} and a
second round of automated flagging was performed using the AUTOFLAG
and MEDNFLAG tasks from Obit. AUTOFLAG clips visibilities with
excessive stokes I and V amplitudes, the latter of which is especially 
useful as most terrestrial interference is highly polarised. ÊMEDNFLAG operates on a source-by-source basis, flagging data whose amplitude is more
discrepant than a given number of sigma times the median amplitude of
the data calculated in a sliding chronological window.

Calibrated data were imaged and cleaned using the Obit IMAGER task, which
has several advantages over the standard AIPS IMAGR. It
implements an algorithm to detect emission and automatically defines
clean windows, leading to a significant reduction in the clean
bias\footnote{See EVLA memo 116
ftp://ftp.cv.nrao.edu/NRAO-staff/bcotton/Obit/autoWindow.pdf.}. In
addition, the task regrids and re-images portions of the image
containing bright compact emission. By centering unresolved sources on
a pixel IMAGER can accurately represent the emission by a single delta
function whilst CLEANing, allowing the creation of images with high
dynamic range\footnote{See EVLA memo 114
ftp://ftp.cv.nrao.edu/NRAO-staff/bcotton/Obit/autoCen.pdf.}. A region
equal to the size of the primary beam (i.e., 4.4$^{\prime}$) was imaged using a
pixel size of 0.3\arcsec, which was chosen to provide $\sim$3 pixels
across the synthesised beam and resulting in image sizes of
1760$\times$1760 pixels.

These maps were then deconvolved using a robust weighting of 0 and up
to a couple of thousand cleaning components -- images were CLEANed down to approximately 3$\sigma$. Finally the processed images were corrected for the attenuation of the primary beam. Due to the nature of interferometric observations the largest well-imaged structure possible at 6~cm, given the limited \emph{uv} coverage and integration time, is $\sim$20$^{\prime\prime}$. However, many of the maps displayed evidence of large-scale emission, which,
when under-sampled, can distort the processed images by producing
image artifacts. These artifacts can appear as large undulations in
the image intensity which are hard to remove and make identification
of weak point sources embedded within extended regions of emission
difficult (see \citealt{vla1999}, p.~127). Alternatively, the
large-scale emission can become over-resolved and break up into
irregular, or multiple component, sources that can often be confused
with real sources. In order to limit the influence of large-scale
emission, these fields were re-imaged, excluding the shortest
baselines, which are most sensitive to extended emission from the
largest angular scales.

\begin{figure}
\begin{center}
\includegraphics[width=0.45\textwidth]{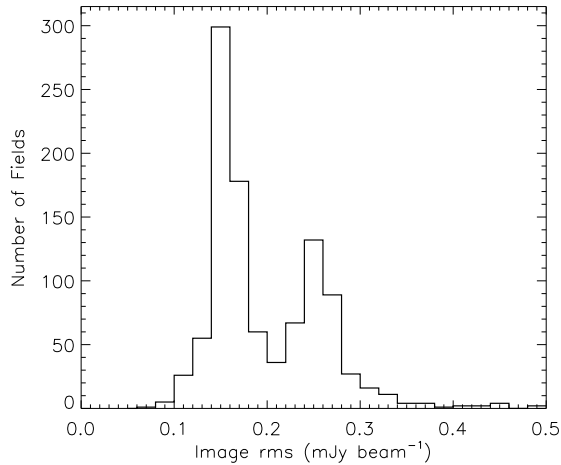}
\caption{\label{fig:field_rms_hist} Histogram of the number of fields observed as a function of map r.m.s. noise level. The double peaked distribution is a result of a smaller bandwidth used for the first season. The data has been binned using a value of 0.02 mJy beam$^{-1}$ wide bins.}

\end{center}
\end{figure}

In Fig.~\ref{fig:field_rms_hist} we present a histogram of the number of observed fields as a function of their r.m.s. noise. The noise levels have been 
estimated from emission free regions close to the centre of the reduced images. This figure clearly shows the effect of the different frequency bandpasses used
for these observations. We have truncated the x-axis at 0.5~mJy
beam$^{-1}$, however, less than 5\% of fields have noise values higher
than this. Typical noise values are $\sim$0.2~mJy beam$^{-1}$, which compares well with the theoretical values. The synthesised beam parameters of the reduced
maps and an estimate of the maps r.m.s. noise level are presented in
the last two columns of Table~3.

\subsection{Source extraction and catalogue}

The reduced maps were examined for compact, high surface brightness
sources using a nominal 4$\sigma$ detection threshold, where $\sigma$
refers to the source's local r.m.s. noise level. Each source found with
a peak flux above the detection threshold was visually inspected in
order to distinguish between genuine sources and possible imaging
artifacts, such as contamination due to bright sidelobes, which were
removed from the final catalogue.

In total we have identified 669 discrete radio sources located within the 1068
fields imaged, although not all necessarily at the position of the RMS source. We name these sources using their Galactic coordinates. In Table~4\footnote{Table~4 and Fig. 2 are only available in electronic form at the CDS via anonymous ftp to cdsarc.u-strasbg.fr (130.79.125.5) or via http://cdsweb.u-strasbg.fr/cgi-bin/qcat?J/A+A/.} we present the
observational parameters for each radio source, i.e., position, peak
and integrated flux, and sizes of the sources' major and minor
axis. In Fig.~2$^7$ we present continuum maps of all
radio sources detected; these are presented in order of increasing
Galactic longitude, are centred on the position of the radio source
given in Table~4 and cover a region of 1
arcmin$^2$ in size. The radio source name is given above each plot and
the size and shape of the restoring beam is indicated by the
ellipse in the lower left corner.

Inspection of the emission maps reveal a variety of different types of morphologies. These are generally classified as: unresolved, Gaussian, spherical, core-halo, cometary, irregular and bipolar (see \citealt{churchwell2002} for a review). Given the observed range of morphological structures a single method for extracting physical parameters is not feasible and therefore a
number of methods have been employed. For sources with a fairly simple
geometry (e.g., unresolved, Gaussian or spherical) we determine their
positions, peak fluxes and sizes using the Miriad task \emph{IMFIT}
which fits a two component Gaussian fit to each radio
source.\footnote{Miriad is a radio interferometry data reduction
package. For more information see
http://www.atnf.csiro.au/computing/software/miriad/.} The sizes given
in Table~4 are the deconvolved sizes
assuming $\theta_{\rm{s}}^2=\theta_{\rm{m}}^2-\theta_{\rm{b}}^2$ where
the subscripted $\theta_{\rm{s}}$, $\theta_{\rm{m}}$ and
$\theta_{\rm{b}}$ refer to the deconvolved full-width at half-maximum
(FWHM) size, the measured FWHM of the source, and the FWHM of the
restoring beam respectively. Since the sizes of the unresolved sources (i.e., sources less than
10\% larger than the beam) are very uncertain we have set an upper
limit to their size which is half the FWHM of the restoring
beam. Furthermore, the unresolved sources tend to be very weak and are
not particularly well fit by the two component Gaussian, leading to
large uncertainties in the integrated fluxes; we have therefore set
the integrated fluxes to be same as the peak fluxes for the unresolved
sources.

For all other morphological types the peak and integrated fluxes have
been determined by summing the flux within a carefully fitted polygon
around each source. For sources classified as having a shell-like
morphology their position is given as the centre of the shell and the
minimum and maximum correspond to the inner and outer diameters of the
shell as measured at the half power points, for core-halo morphologies the
major and minor axes of the core have been given. For sources with
extended, irregular or cometary morphologies their sizes have been
determined by eye. As all of these morphological types are extended
with respect to the beam their sizes have not been deconvolved with
the beam.

The radio sources detected consist of two distinct populations; thermal emission from Galactic sources such as HII regions and PNe, and an extragalactic population of non-thermal emission (e.g., active-galactic nuclei). Cross matching these radio sources with the MSX point source catalogue we find approximately a third are associated with mid-infrared emission and therefore likely to be Galactic in origin. (In the last two columns of Table~4 we present the positional offset between the radio and MSX positions and the MSX point source name.) Two-thirds of the radio sources detected have no thermal counterpart. These radio sources maybe associated with with a relatively weak thermal source not detected by MSX. \citet{giveon2005a} estimated the MSX catalogue to be $\geq$80\% complete for thermal sources, and therefore the majority of the unassociated radio sources are like to be extragalactic in origin. 

\begin{figure*}
\begin{center}

\includegraphics[width=0.32\linewidth, trim=0 0 25 0]{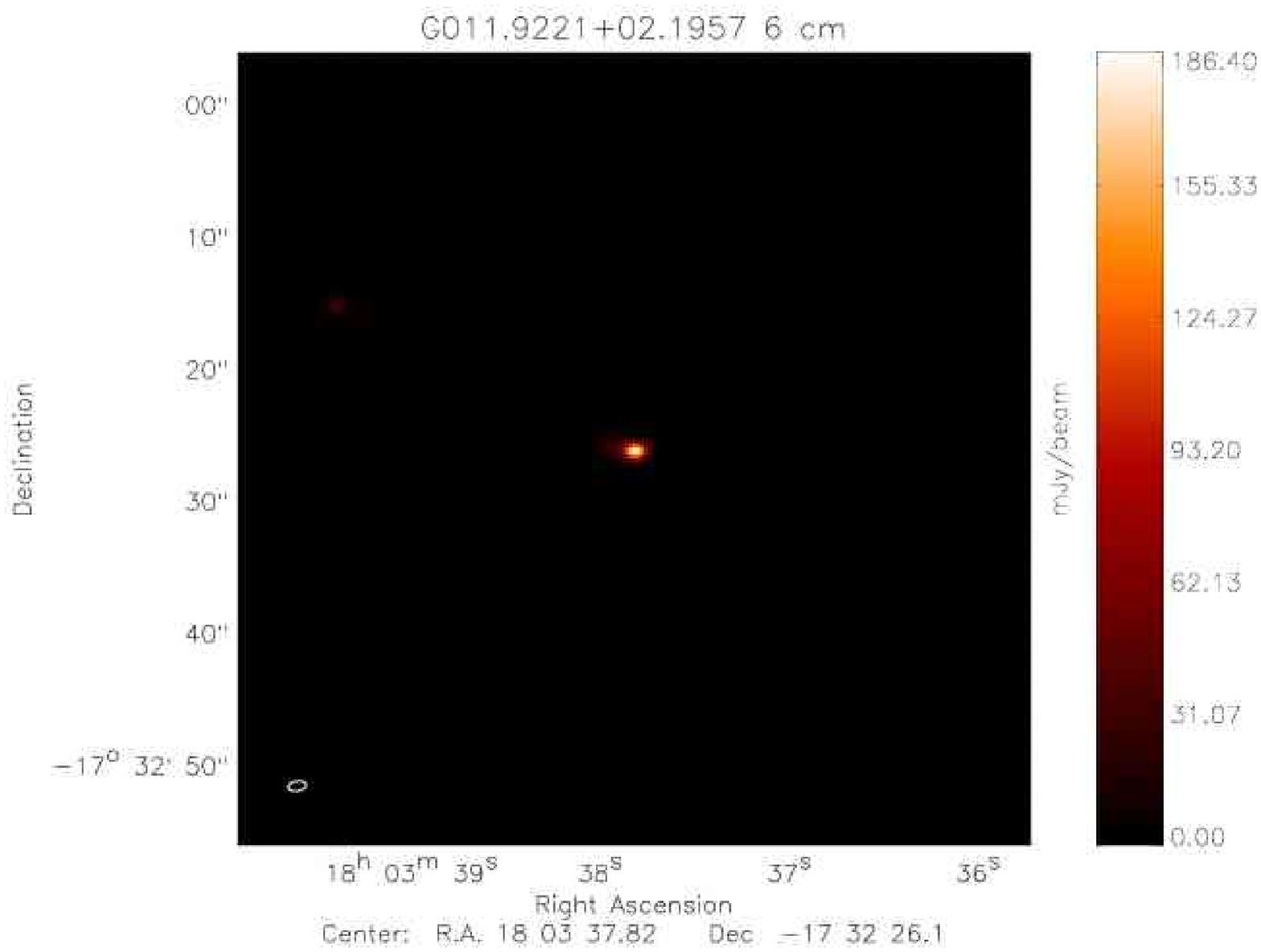}
\includegraphics[width=0.32\linewidth, trim=0 0 25 0]{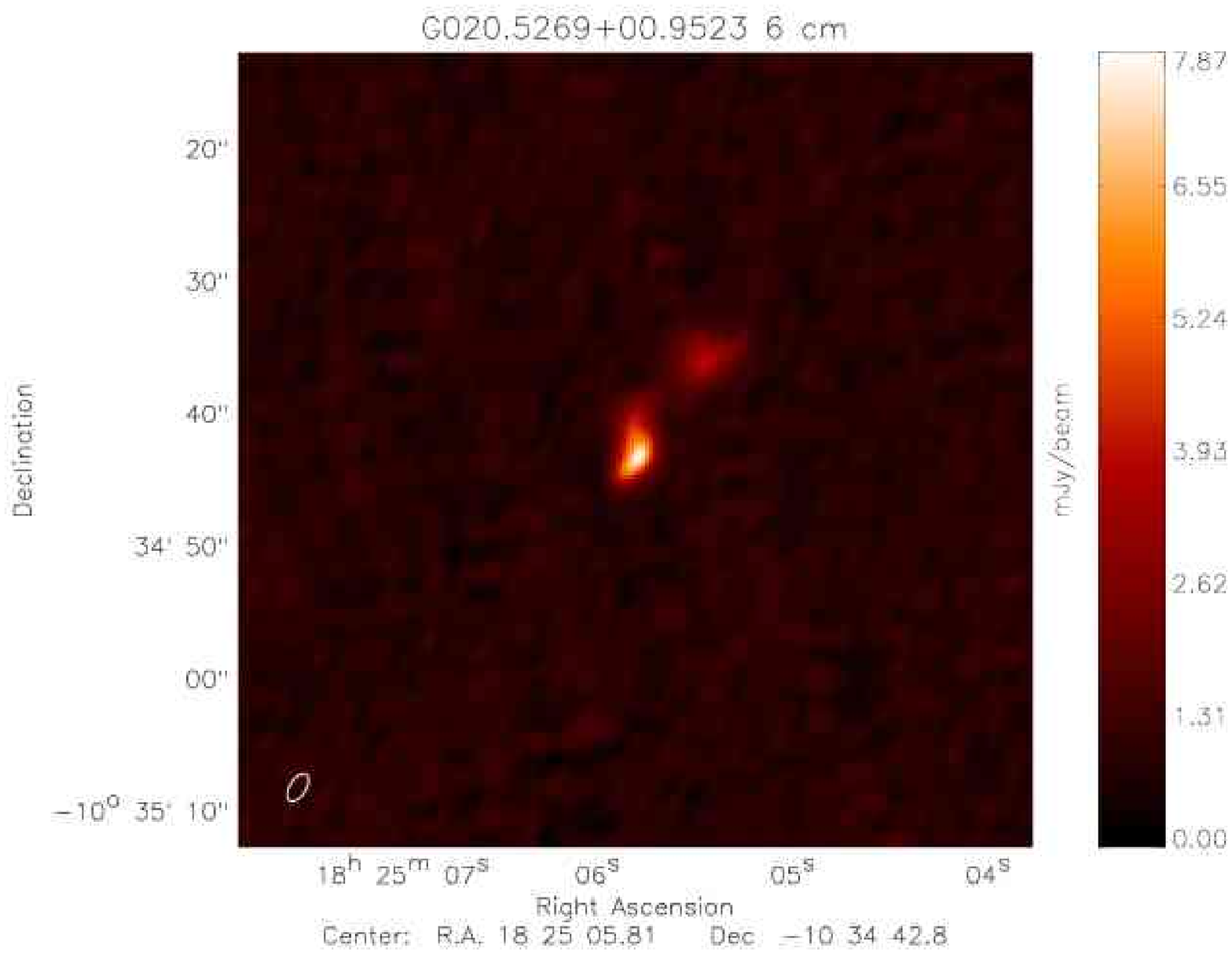}
\includegraphics[width=0.32\linewidth, trim=0 0 25 0]{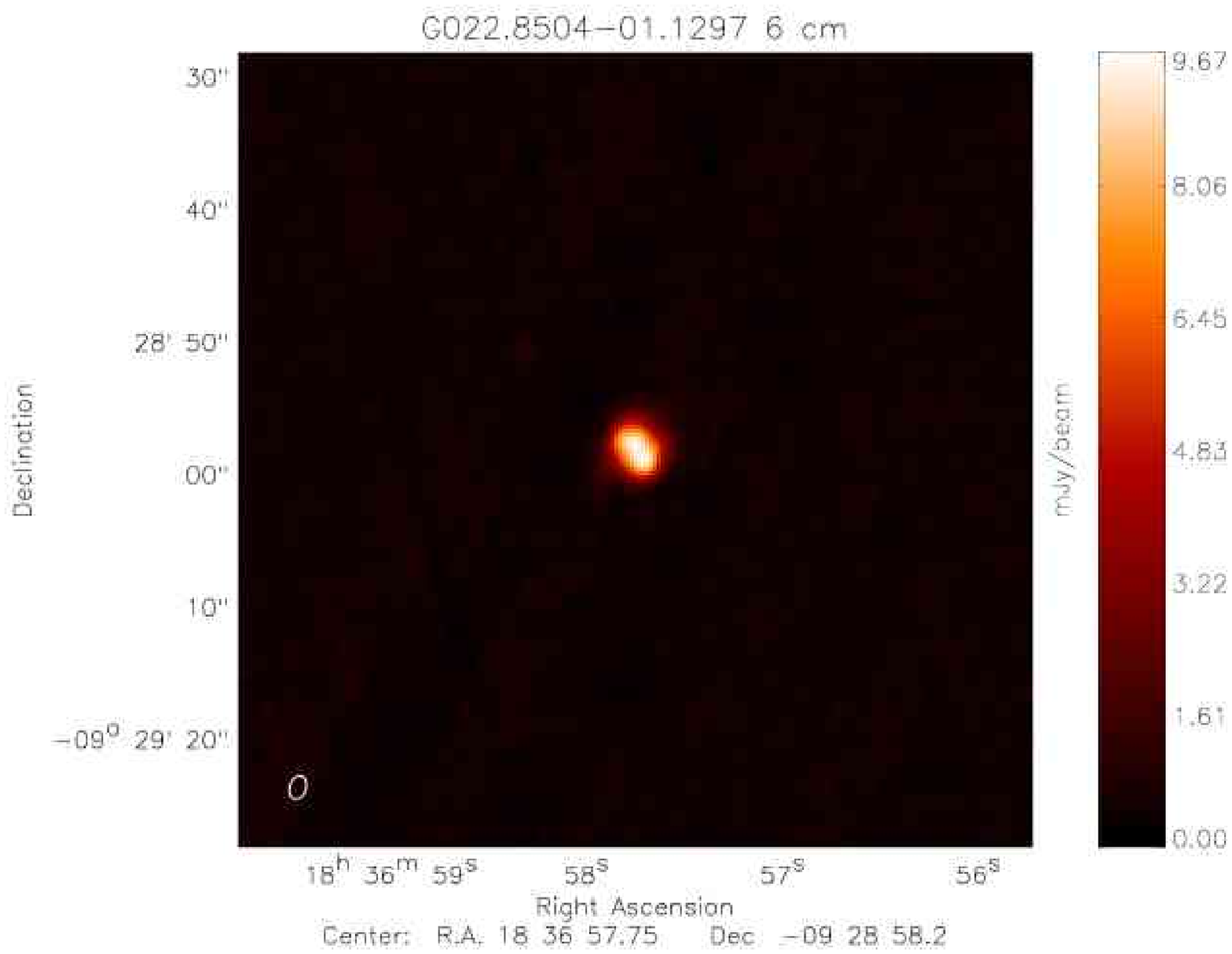}

\includegraphics[width=0.32\linewidth, trim=0 0 25 0]{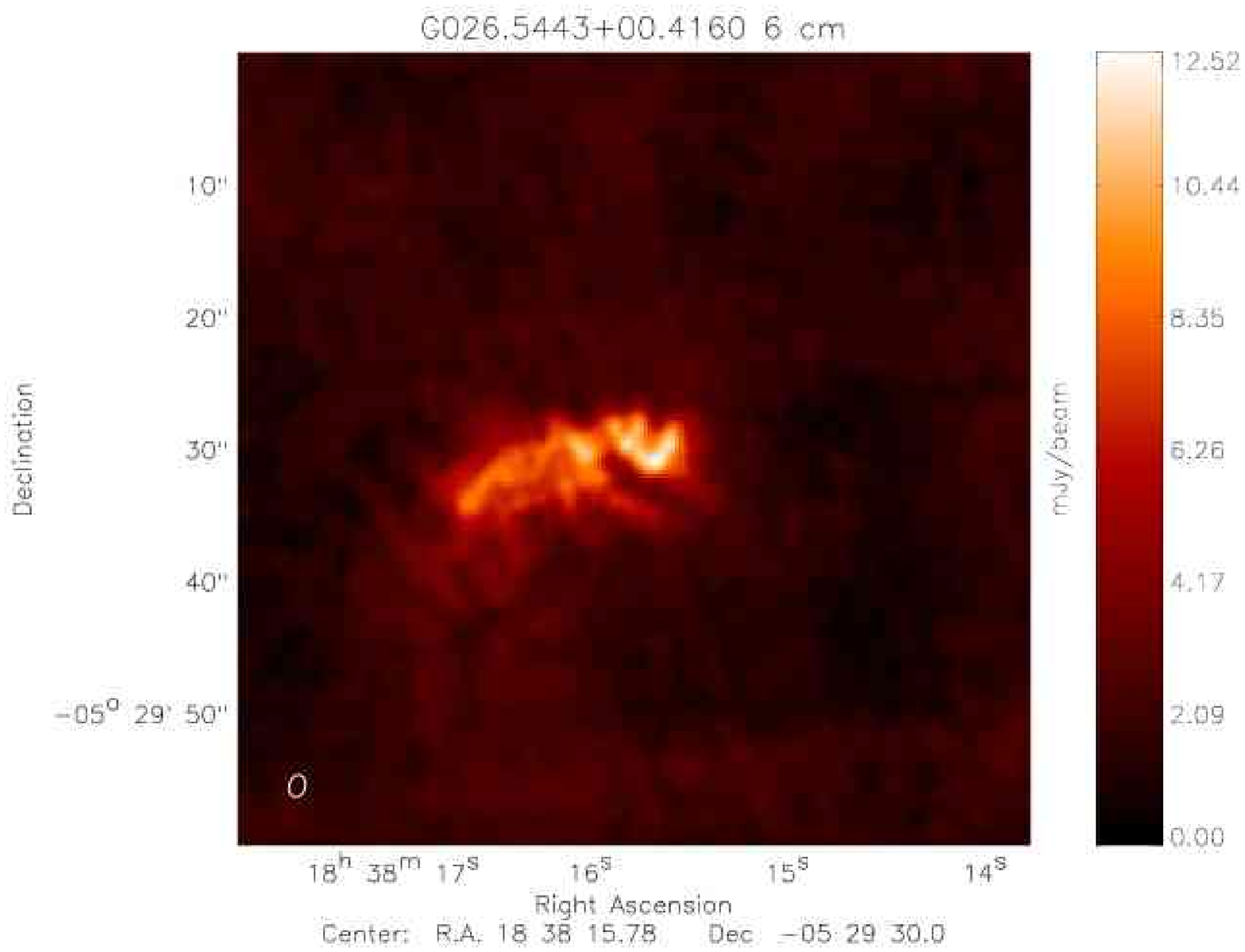}
\includegraphics[width=0.32\linewidth, trim=0 0 25 0]{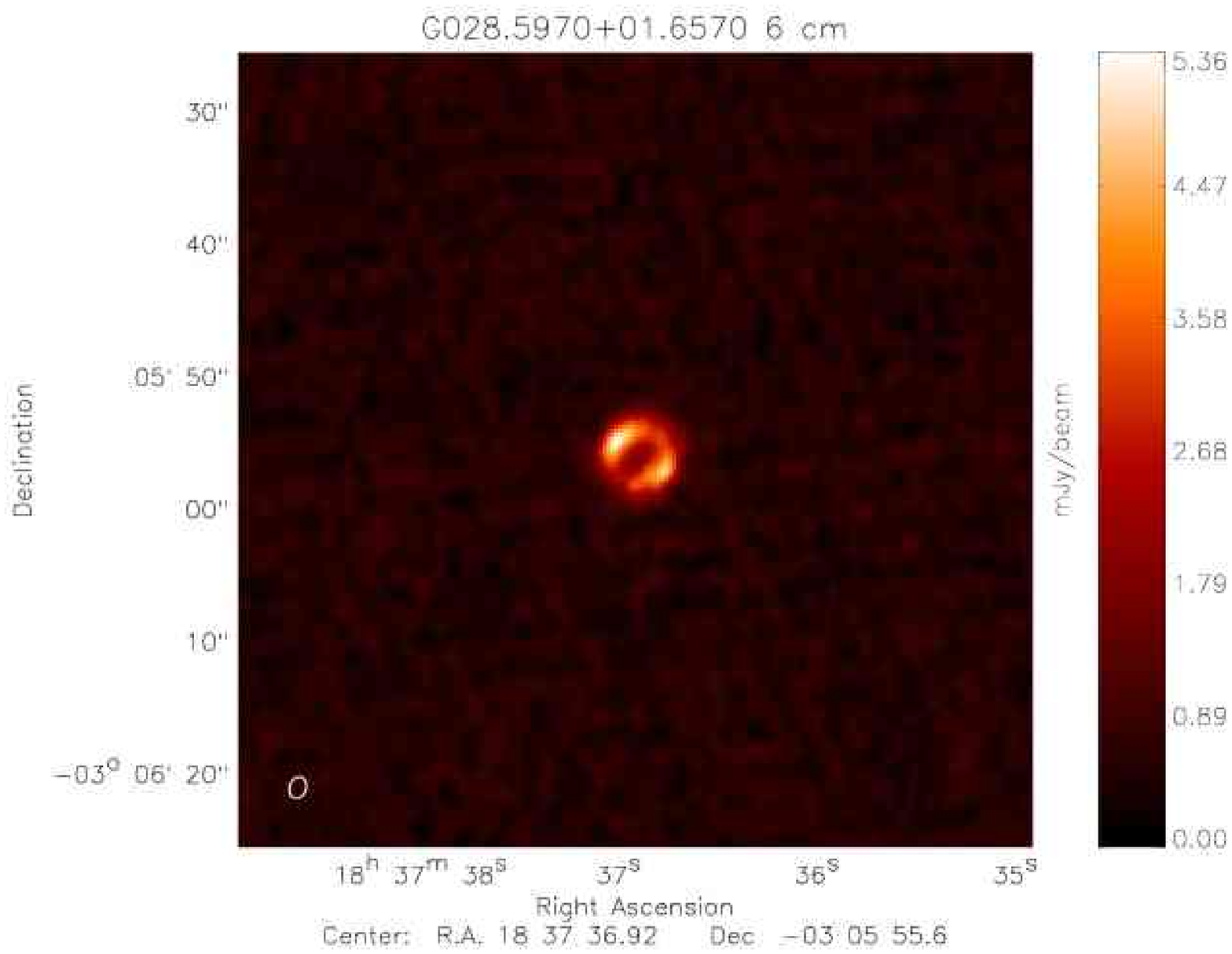}
\includegraphics[width=0.32\linewidth, trim=0 0 25 0]{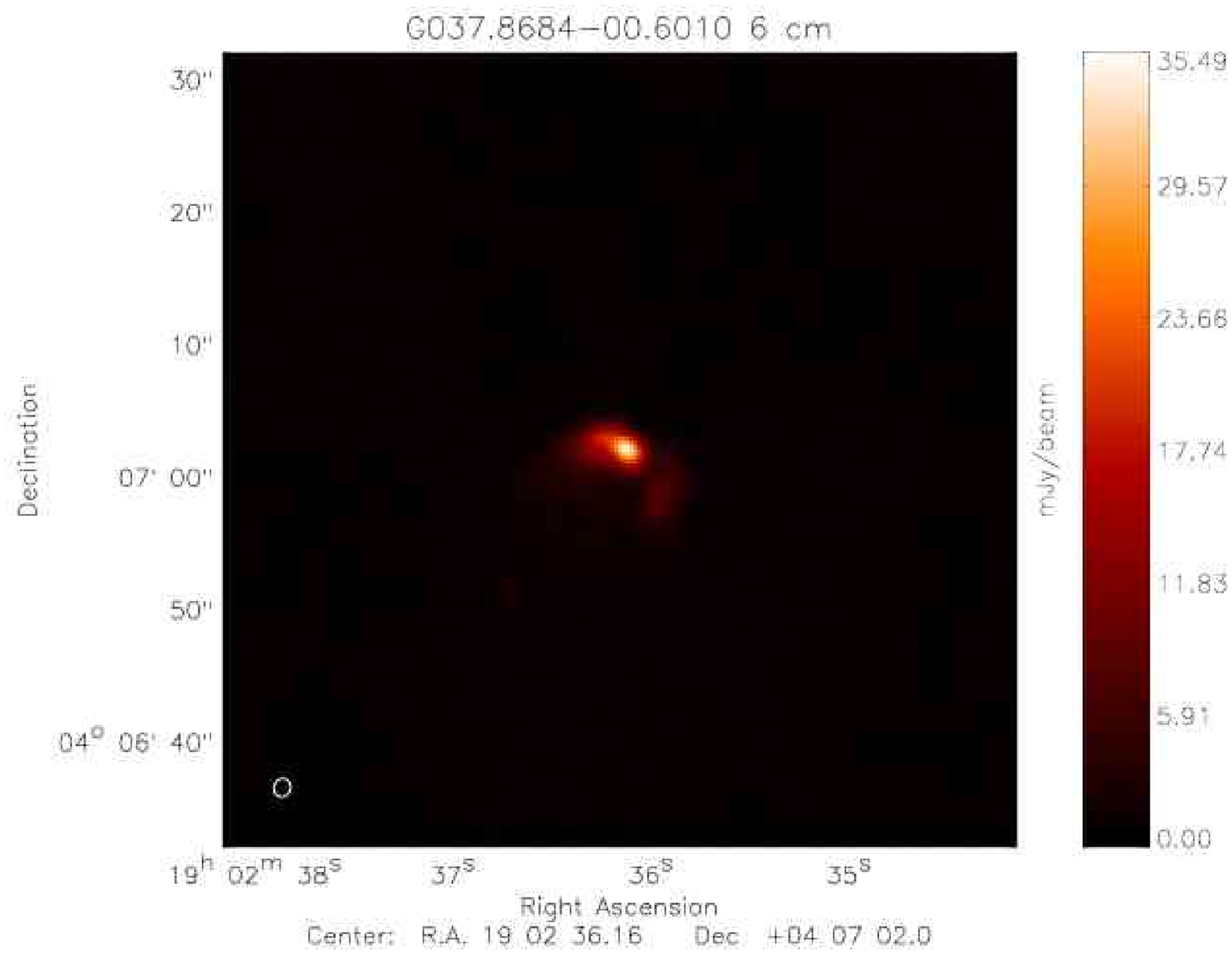}

\includegraphics[width=0.32\linewidth, trim=0 0 25 0]{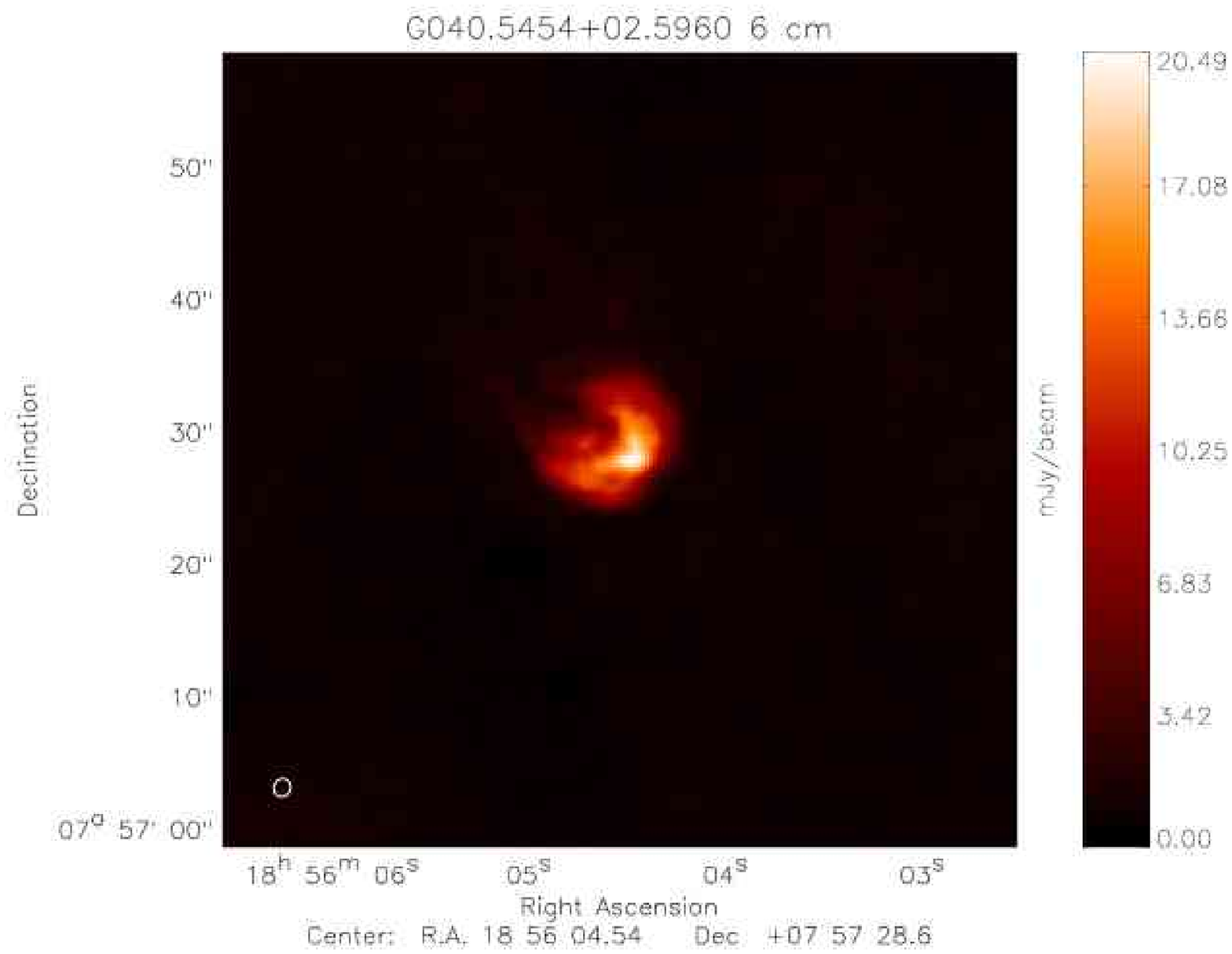}
\includegraphics[width=0.32\linewidth, trim=0 0 25 0]{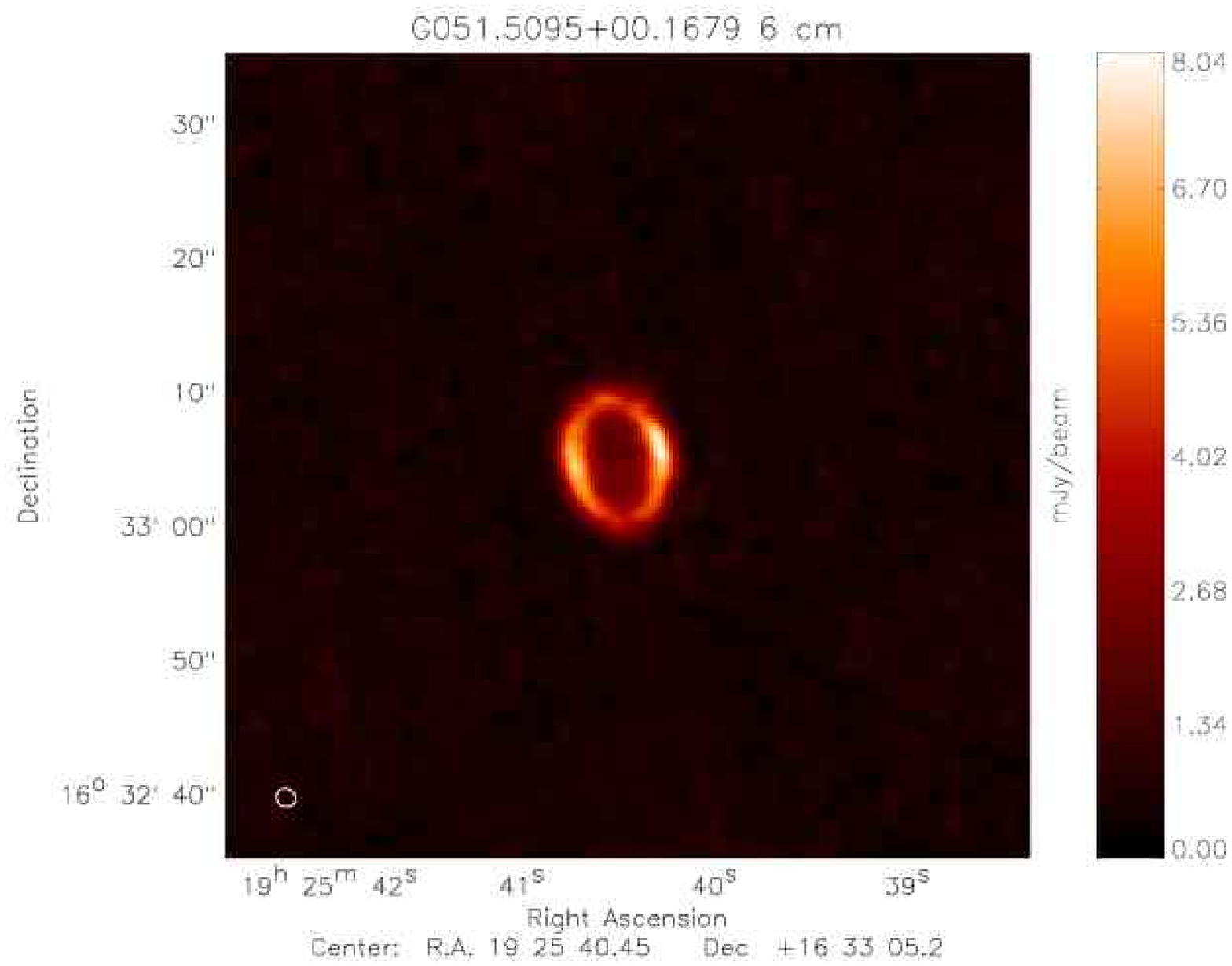}
\includegraphics[width=0.32\linewidth, trim=0 0 25 0]{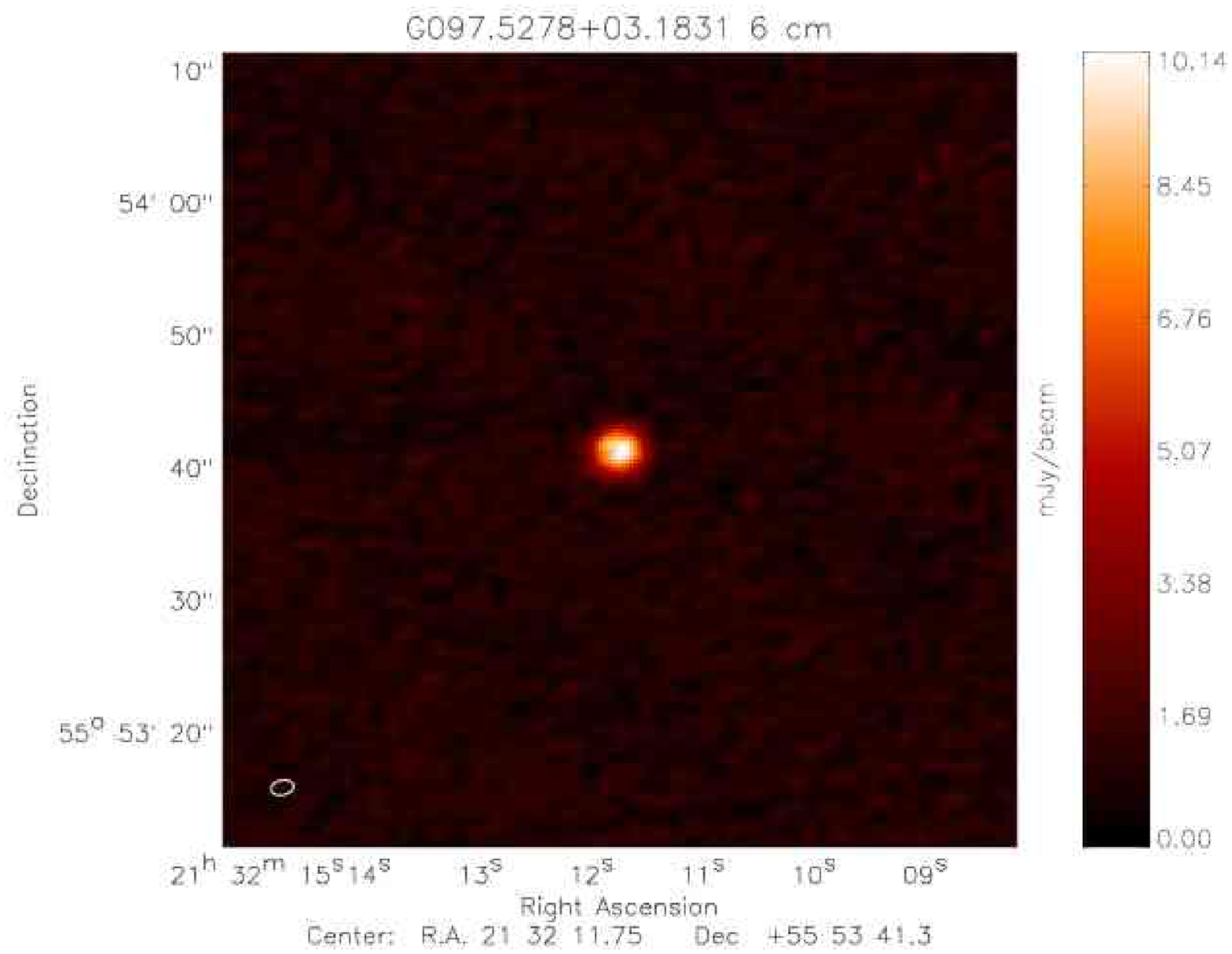}

\caption{\label{fig:radio_maps} Images (1\arcmin$\times$1\arcmin) of all radio sources detected by these targeted VLA observations. The source name is printed above each image and shape and orientation of the restoring beam is shown in the lower left corner. Only a small sample of images are provided here the full version if this figure is available in electronic form at the CDS via anonymous ftp to cdsarc.u-strasbg.fr (130.79.125.5) or via http://cdsweb.u-strasbg.fr/cgi-bin/qcat?J/A+A/.}
\end{center}
\end{figure*}

\begin{table*}
\begin{center}
\caption{VLA radio detection catalogue. }
\label{tbl:detection_parameters}
\begin{minipage}{\linewidth}
\begin{tabular}{lccrrrlcc}
\hline
\hline
Radio Name	&	RA	&	Dec.	&	\multicolumn{1}{c}{Peak} 	&	\multicolumn{1}{c}{Integrated}	&	&Source	&	\multicolumn{1}{c}{MSX Offset}&	MSX Name	\\
	&	(J2000)	&	(J2000)	&	\multicolumn{1}{c}{(mJy)} 	&	\multicolumn{1}{c}{(mJy)}	&	&Size (\arcsec)	&	 \multicolumn{1}{c}{(\arcsec)}	&		\\

\hline
G010.0653$-$02.0583	&	18:15:36.97	&	$-$21:13:21.2	&	2.5	&	10.8	&		&	3.2 $\times$ 2.3	&		&		\\
G010.3924+00.5390	&	18:06:34.26	&	$-$19:41:05.3	&	0.7	&	2.2	&		&	4.6 $\times$ 2.1	&	2.3	&	G010.3930+00.5389	\\
G010.5959$-$00.8734	&	18:12:14.95	&	$-$20:11:25.1	&	11.3	&	59.4	&		&	1.5 $\times$ 1.4	&	2.9	&	G010.5956$-$00.8740	\\
G010.5971$-$00.3893	&	18:10:26.69	&	$-$19:57:22.4	&	8.2	&	22.5	&		&	2.3 $\times$ 1.4	&		&		\\
G010.6003$-$00.8320	&	18:12:06.21	&	$-$20:09:59.8	&	1.8	&	3.2	&		&	0.8 $\times$ 0.4	&		&		\\
G010.6161$-$00.3921	&	18:10:29.65	&	$-$19:56:27.5	&	32.8	&	99.1	&		&	2.4 $\times$ 1.5	&		&		\\
G010.6221$-$00.3788	&	18:10:27.42	&	$-$19:55:45.3	&	42.6	&	104.1	&		&	2.8 $\times$ 0.8	&		&		\\
G010.6234$-$00.3838	&	18:10:28.69	&	$-$19:55:50.1	&	212.2	&	1055.0	&		&	3.3 $\times$ 2.2	&	1.5	&	G010.6235$-$00.3834	\\
G010.6240$-$00.3812	&	18:10:28.19	&	$-$19:55:43.6	&	33.6	&	86.6	&		&	1.9 $\times$ 1.6	&	8.3	&	G010.6235$-$00.3834	\\
G011.1073+00.4272	&	18:08:27.60	&	$-$19:06:52.5	&	1.2	&	1.6	&		&	1.7 $\times$ 0.6	&		&		\\
G011.4201$-$01.6819	&	18:16:56.89	&	$-$19:51:08.1	&	0.8	&	3.2	&		&	1.9 $\times$ 1.5	&	1.1	&	G011.4201$-$01.6815	\\
G011.6218+01.1804	&	18:06:44.19	&	$-$18:17:56.0	&	4.3	&	4.7	&		&	0.5 $\times$ 0.4	&		&		\\
G011.8724+02.2273	&	18:03:24.77	&	$-$17:34:06.0	&	3.8	&	4.9	&		&	0.9 $\times$ 0.3	&		&		\\
G011.9221+02.1957	&	18:03:37.82	&	$-$17:32:26.2	&	189.2	&	340.9	&		&	1.0 $\times$ 0.7	&		&		\\
G011.9278+02.1917	&	18:03:39.40	&	$-$17:32:15.1	&	26.7	&	45.2	&		&	1.0 $\times$ 0.4	&		&		\\
G012.1098+01.2250	&	18:07:34.31	&	$-$17:51:03.3	&	3.3	&	4.6	&		&	1.9 $\times$ 0.8	&		&		\\
G012.1175+01.1965	&	18:07:41.54	&	$-$17:51:28.9	&	32.2	&	63.6	&		&	1.9 $\times$ 1.5	&	1.2	&	G012.1177+01.1966	\\
G012.4180+00.5038	&	18:10:51.08	&	$-$17:55:49.5	&	3.5	&	6.2	&		&	1.4 $\times$ 1.2	&		&		\\
G012.4317$-$01.1112	&	18:16:51.23	&	$-$18:41:28.4	&	8.1	&	105.2	&		&	3.7 $\times$ 3.1	&	2.2	&	G012.4314$-$01.1117	\\
G012.8162+00.5576	&	18:11:27.53	&	$-$17:33:20.1	&	7.7	&	9.3	&		&	0.5 $\times$ 0.1	&		&		\\
G012.8616+00.8776	&	18:10:22.50	&	$-$17:21:41.5	&	1.6	&	1.8	&		&	0.8 $\times$ 0.2	&		&		\\
G012.9630+00.9441	&	18:10:20.15	&	$-$17:14:26.2	&	1.2	&	1.2	&	$<$	&	2.2 $\times$ 0.6	&		&		\\
G012.9712+00.9055	&	18:10:29.63	&	$-$17:15:07.4	&	1.3	&	1.3	&	$<$	&	2.2 $\times$ 0.6	&	10	&	G012.9731+00.9036	\\
G013.1178+04.1527	&	17:58:58.77	&	$-$15:32:14.3	&	30.2	&	70.2	&		&	2.2 $\times$ 2.0	&	7.9	&	G013.1172+04.1507	\\
G013.4895$-$00.7408	&	18:17:35.67	&	$-$17:35:04.0	&	2.9	&	5.1	&		&	1.9 $\times$ 0.8	&		&		\\
\hline
\end{tabular}\\
\end{minipage}
\end{center}
Notes: Only a small portion of the data is provided here, the full table is available in electronic form at the CDS via anonymous ftp to cdsarc.u-strasbg.fr (130.79.125.5) or via http://cdsweb.u-strasbg.fr/cgi-bin/qcat?J/A+A/.

\end{table*}

\setcounter{figure}{2}
\setcounter{table}{4}

It is the radio detections associated with the RMS sources that are
of most interest to us and which will be the focus for the rest of this paper. The other radio detections will not be discussed further, however,
we present their measured parameter and emission maps in Table~4 and
Fig.~2 respectively as they may prove useful to the wider
community.

\section{Identifying RMS-radio associations}

\subsection{VLA observations}
\label{sect:vla_obs}

Within these fields are located 1125 sources which passed our initial
MSX colour selection, however, 466 were later excluded either because
they were subsequently found to be associated with 2MASS sources that
possess flat, or blue, near-infrared colours, or were found to be
considerably extended rather than point sources in the MSX images and
therefore more likely to be HII~regions than MYSOs. Therefore the VLA
observations presented here provide radio continuum data for the 659 RMS
sources that remain from the original RMS catalogue of $\sim$2000 MYSO
candidates.  In order to identify radio sources that are associated
with our RMS sources we compared the positions of the radio detections
with the positions of the 659 RMS sources within the observed
fields. We found a total of 167 radio sources within an initially generous  25\arcsec\ search radius of 143 RMS sources.

\subsection{Archival data}
\label{sect:magpis}

As previously mentioned these targeted observations were tailored to avoid a large region of the
Galactic plane that had previously been surveyed by \citet{becker1994}
with the VLA. This survey covers the inner Galaxy (350\degr\ $< l < $
42\degr\, $|b|$ $<$ 0.4\degr) with a resolution of $\sim$6\arcsec\
(9\arcsec$\times$4\arcsec\ restoring beam) and a median sensitivity of
$\sim$0.18~mJy, however, similarly to the targeted observations, approximately
5\% of their fields have noise levels that exceed this value by more
than a factor of two. This survey was recently reprocessed using
improved phase calibration and incorporating some re-observed fields
(\citealt{white2005}). These improvements resulted in a more uniform
survey and an almost three fold increase in the number of sources
detected from 1272 to 3283.

We searched the \citet{white2005} catalogue for possible radio matches
for the 315 RMS sources that are located within this survey
region. This search identified $\sim$250 radio sources within a
25\arcsec\ of 149 RMS sources. In order to compare the spatial distribution of
the potential radio sources matched with the RMS sources we obtained
images from the The Multi-Array Galactic Plane Imaging Survey (MAGPIS)
project page.\footnote{http://third.ucllnl.org/gps/}

\subsection{Selection criterion}
\label{sect:selection}

\begin{figure}
\begin{center}

\includegraphics[width=0.45\textwidth]{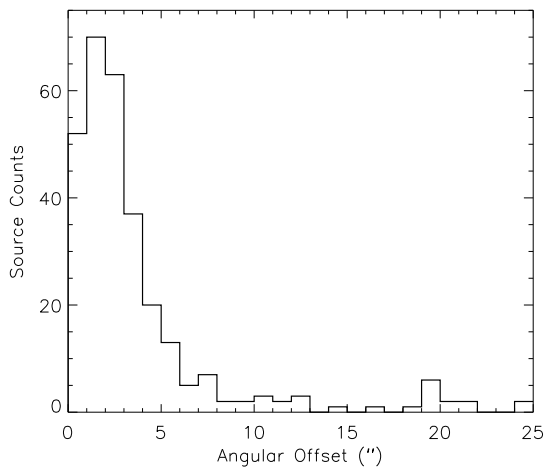}

\caption{\label{fig:rms_radio_offsets} Histogram of the projected angular separations between the RMS sources and their nearest radio source. The radio sources have been drawn from the targeted VLA observation and those found in the \citet{white2005} catalogue. The data have been binned using a value of 1\arcsec.}

\end{center}
\end{figure}

In the previous two subsections we used a rather generous search
radius of 25\arcsec\ to identify possible RMS-radio associations. Using
this radius we identifed a total of 417 radio sources positionally
coincident with 292 RMS sources. However, due to the size of the
search radius there is a non-negligible possibility of chance
alignments due to line of sight contamination, and therefore not all of
these are likely to be genuine associations. In order to try to
distinguish between real and chance associations we examined the angular correlation between the RMS and radio sources to derive an association criterion. We present a histogram plot of the angular separation between the radio sources and their nearest RMS source in Fig.~\ref{fig:rms_radio_offsets}.

The distribution of the angular separation between the radio and RMS
matches illustrates the tight correlation between the two at
small angular offsets. The distribution is strongly peaked at
$\sim$1--3\arcsec\ with typical separations of $\sim$2\arcsec, which
is comparable to the positional accuracy of the MSX point source
catalogue ($\sim$2\arcsec; \citealt{egan1999}). After peaking sharply
the distribution falls off steeply to $\sim$6\arcsec\ and then more
slowly until $\sim$13\arcsec, after which the distribution tails off
completely. This distribution closely matches the distribution found
towards a sample of southern RMS-radio associations observed with the
ATCA (cf. Fig.~2 from Paper~I).

We consider the point at which the distribution tails off to mark the
transition between real associations and chance alignments. The
distribution begins to flatten out between 8\arcsec\ and 15\arcsec\
and the selection of a cutoff point between these values is quite
arbitrary. We have therefore selected a radius of 12\arcsec\ for
reliable associations of radio and RMS sources. We choose this value because it  is the mid-point between the tail of the distribution and for consistency with the cutoff used for the ATCA observations (Paper~I).

Applying this radial cutoff we find over 90\% of the matched sources
are located within 12\arcsec\ of each other, and more than 85\% of
these are located within a 5\arcsec\ radius. In order to check whether our
cutoff radius was giving reliable associations we followed the
procedures described by \citet{giveon2005a} for identifying
radio-infrared matches. This analysis resulted in a reliability of
97\% or better for every RMS source matched with a radio source within
a 12\arcsec\ radius.

\section{Results and Analysis}

\subsection{RMS-radio associations}

In total 974 RMS sources are located within either the VLA fields
observed as part of our programme of targeted
observations or lie within the region of the plane reported by \citet{white2005} (see Sect.~\ref{sect:magpis}).  Searching the catalogues associated with these two programmes we find radio emission associated with 272 of the 974 RMS sources for which data is availablle. These associations amount to $\sim$27\% of the RMS sample observed (hereafter we will refer to these as RMS-radio matches). In about 20\% of case we find two or more radio sources are found to be associated with a single RMS source; this brings the total number of radio source that lie within a 12\arcsec\ radius (see previous section) of an RMS source to 342. In Table~\ref{tbl:rms_radio} we present the MSX names of the sources, the radio name, the angular offset, peak and integrated fluxes, angular sizes and source classifications (see Sect.~\ref{sect:classification} for discussion concerning source classification).


\renewcommand{\thefootnote}{\alph{footnote}}
\setcounter{footnote}{0}
\begin{table*}
\begin{minipage}{\linewidth}
\begin{center}
\caption{Parameters for radio sources found to be associated with RMS sources. }
\label{tbl:rms_radio}

\begin{tabular}{llcrrrrlc}
\hline
\hline
RMS Name &Radio Name\footnotemark \setcounter{footnote}{0}	& Offset &	RA	&	Dec.	&	Peak 	&	Integrated	&	Source	&	Classification\footnotemark \setcounter{footnote}{2} 	\\
& & (\arcsec)	&	(J2000)	&	(J2000)	&	(mJy) 	&	(mJy)	&	Size (\arcsec)	&		\\
\hline
G010.4413+00.0101	&	G10.440+0.011$^\dagger$	&	4.4	&	18:08:37.93	&	$-$19:53:58.9	&	7.5	&	32.5	&	13.1 $\times$ 8.1	&	HII region	\\
G010.5975$-$00.3838	&	G10.598$-$0.384$^\dagger$	&	0.6	&	18:10:25.59	&	$-$19:57:11.0	&	66.1	&	136.6	&	7.7 $\times$ 4.7	&	HII region	\\
	&	G10.597$-$0.383$^\dagger$	&	6.0	&	18:10:25.13	&	$-$19:57:12.2	&	52.0	&	105.6	&	12.6 $\times$ 2.4	&	HII region	\\
	&	G10.601$-$0.384$^\dagger$	&	11.8	&	18:10:26.03	&	$-$19:57:01.8	&	32.9	&	78.9	&	12.5 $\times$ 4.0	&	HII region	\\
G010.6291$-$00.3385	&	G10.629$-$0.338$^\dagger$	&	1.2	&	18:10:19.25	&	$-$19:54:12.1	&	18.1	&	25.2	&	4.6 $\times$ 2.9	&	HII region	\\
G010.9592+00.0217	&	G10.959+0.022$^\dagger$	&	3.2	&	18:09:39.35	&	$-$19:26:26.9	&	178.7	&	195.0	&	2.2 $\times$ 1.0	&	HII region	\\
G010.9657+00.0083	&	G10.967+0.009$^\dagger$	&	3.6	&	18:09:43.31	&	$-$19:26:25.1	&	16.4	&	68.7	&	12.3 $\times$ 7.9	&	HII region	\\
	&	G10.965+0.010$^\dagger$	&	5.9	&	18:09:42.95	&	$-$19:26:27.8	&	34.2	&	63.8	&	5.9 $\times$ 4.7	&	HII region	\\
	&	G10.965+0.006$^\dagger$	&	8.3	&	18:09:43.64	&	$-$19:26:36.0	&	19.8	&	74.8	&	13.0 $\times$ 6.6	&	HII region	\\
G011.1109$-$00.4001	&	G11.111$-$0.400$^\dagger$	&	0.3	&	18:11:32.30	&	$-$19:30:40.7	&	18.2	&	50.0	&	9.9 $\times$ 3.4	&	HII region	\\
	&	G11.111$-$0.399$^\dagger$	&	6.0	&	18:11:31.88	&	$-$19:30:38.5	&	96.7	&	111.9	&	3.7 $\times$ 1.2	&	HII region	\\
G011.1723$-$00.0656	&	G11.172$-$0.065$^\dagger$	&	3.4	&	18:10:24.98	&	$-$19:17:46.1	&	23.5	&	71.0	&	9.8 $\times$ 6.9	&	HII region	\\
G011.4201$-$01.6815	&	G011.4201$-$01.6819	&	1.1	&	18:16:56.89	&	$-$19:51:08.1	&	0.8	&	3.2	&	1.9 $\times$ 1.5	&	HII/YSO	\\
G011.9454$-$00.0373	&	G11.946$-$0.036$^\dagger$	&	5.1	&	18:11:53.22	&	$-$18:36:14.6	&	70.2	&	623.4	&	18.3 $\times$ 14.9	&	HII region	\\
	&	G11.944$-$0.037$^\dagger$	&	5.1	&	18:11:53.20	&	$-$18:36:21.8	&	271.6	&	453.4	&	4.7 $\times$ 3.9	&	HII region	\\
G012.1177+01.1966	&	G012.1175+01.1965	&	1.2	&	18:07:41.54	&	$-$17:51:28.9	&	32.2	&	63.6	&	1.9 $\times$ 1.5	&	PN	\\
G012.1917$-$00.1029	&	G12.192$-$0.103$^\dagger$	&	1.0	&	18:12:37.95	&	$-$18:25:15.0	&	21.3	&	119.6	&	18.6 $\times$ 7.2	&	HII region	\\
	&	G12.191$-$0.101$^\dagger$	&	7.3	&	18:12:37.48	&	$-$18:25:13.3	&	9.5	&	30.1	&	12.4 $\times$ 5.7	&	HII region	\\
G012.1993$-$00.0342	&	G12.199$-$0.034$^\dagger$	&	3.0	&	18:12:23.59	&	$-$18:22:54.4	&	54.3	&	60.1	&	1.8 $\times$ 1.3	&	HII/YSO	\\
G012.4314$-$01.1117	&	G012.4317$-$01.1112	&	2.2	&	18:16:51.23	&	$-$18:41:28.4	&	8.1	&	105.2	&	3.7 $\times$ 3.1	&	HII region	\\
G012.8062$-$00.1987	&	G12.805$-$0.200$^\dagger$	&	5.3	&	18:14:13.74	&	$-$17:55:42.1	&	1748.5	&	6290.8	&	12.6 $\times$ 4.6	&	HII region	\\
G013.2097$-$00.1436	&	G13.210$-$0.144$^\dagger$	&	2.4	&	18:14:50.05	&	$-$17:32:44.7	&	317.5	&	1017.0	&	9.1 $\times$ 7.8	&	HII region	\\
G014.1742+00.0226	&	G14.175+0.024$^\dagger$	&	3.9	&	18:16:08.53	&	$-$16:37:06.1	&	21.2	&	96.4	&	11.6 $\times$ 10.0	&	HII region	\\
G014.2071$-$00.1105	&	G14.207$-$0.110$^\dagger$	&	3.3	&	18:16:41.75	&	$-$16:39:12.1	&	11.4	&	60.0	&	13.1 $\times$ 11.1	&	HII region	\\
G014.2361+00.2138	&	G14.237+0.212$^\dagger$	&	7.0	&	18:15:34.45	&	$-$16:28:27.1	&	30.2	&	35.5	&	2.3 $\times$ 2.2	&	PN	\\
\hline
\end{tabular}\\

\footnotetext[1]{Matches identified from the \citealt{white2005} catalogue are identified by a superscript $\dagger$ added to the end of the source's radio name.}
\footnotetext[2]{RMS sources have been give one of five difference classifications: HII region, PN, YSO, HII/YSO and Other. The first three of these are obvious, however, the last two require a little explanation. The HII/YSO classification is give to MSX sources that appear to be associated with both an HII region and a YSO. The classification `Other' is given to sources that do not fit into any of the other classification. }

\end{center}
\end{minipage}
Notes: Only a small portion of the data is provided here, the full table is available in electronic form at the CDS via anonymous ftp to cdsarc.u-strasbg.fr (130.79.125.5) or via http://cdsweb.u-strasbg.fr/cgi-bin/qcat?J/A+A/.

\end{table*}

\renewcommand{\thefootnote}{\arabic{footnote}}
\setcounter{footnote}{9}

\begin{figure*}
\begin{center}
\includegraphics[width=0.32\linewidth, trim=0 0 150 0]{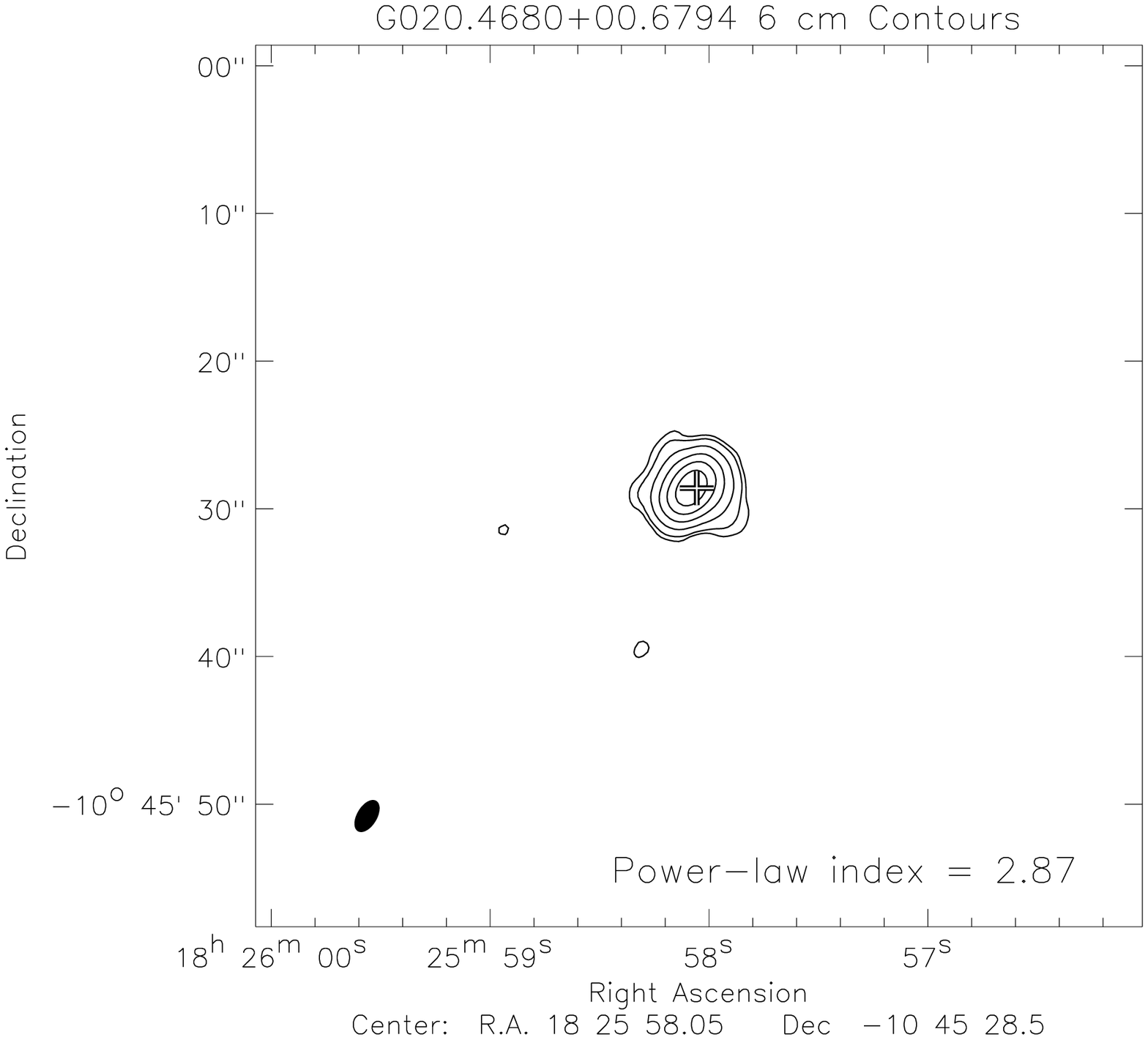}
\includegraphics[width=0.32\linewidth, trim=0 0 150 0]{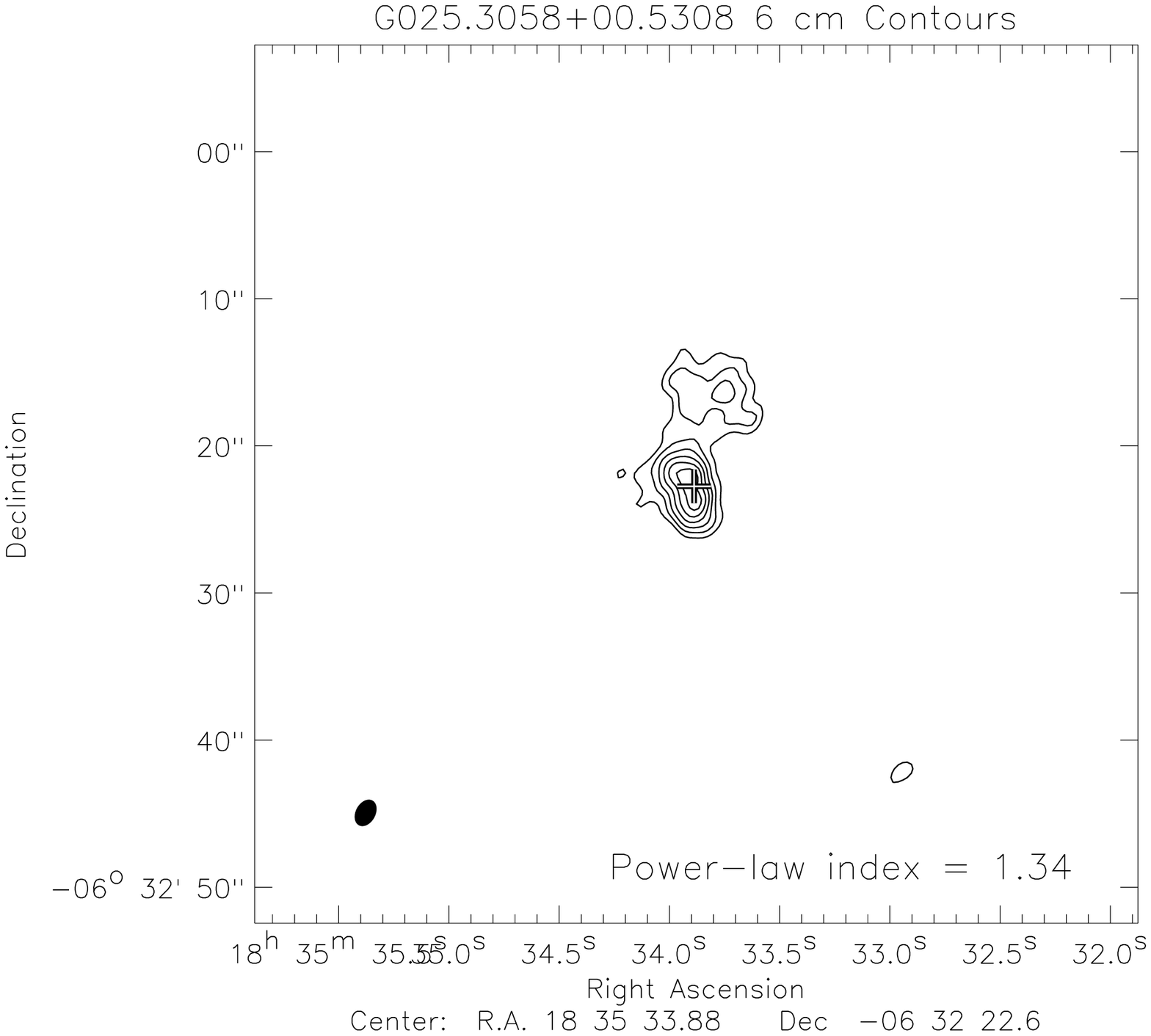}
\includegraphics[width=0.32\linewidth, trim=0 0 150 0]{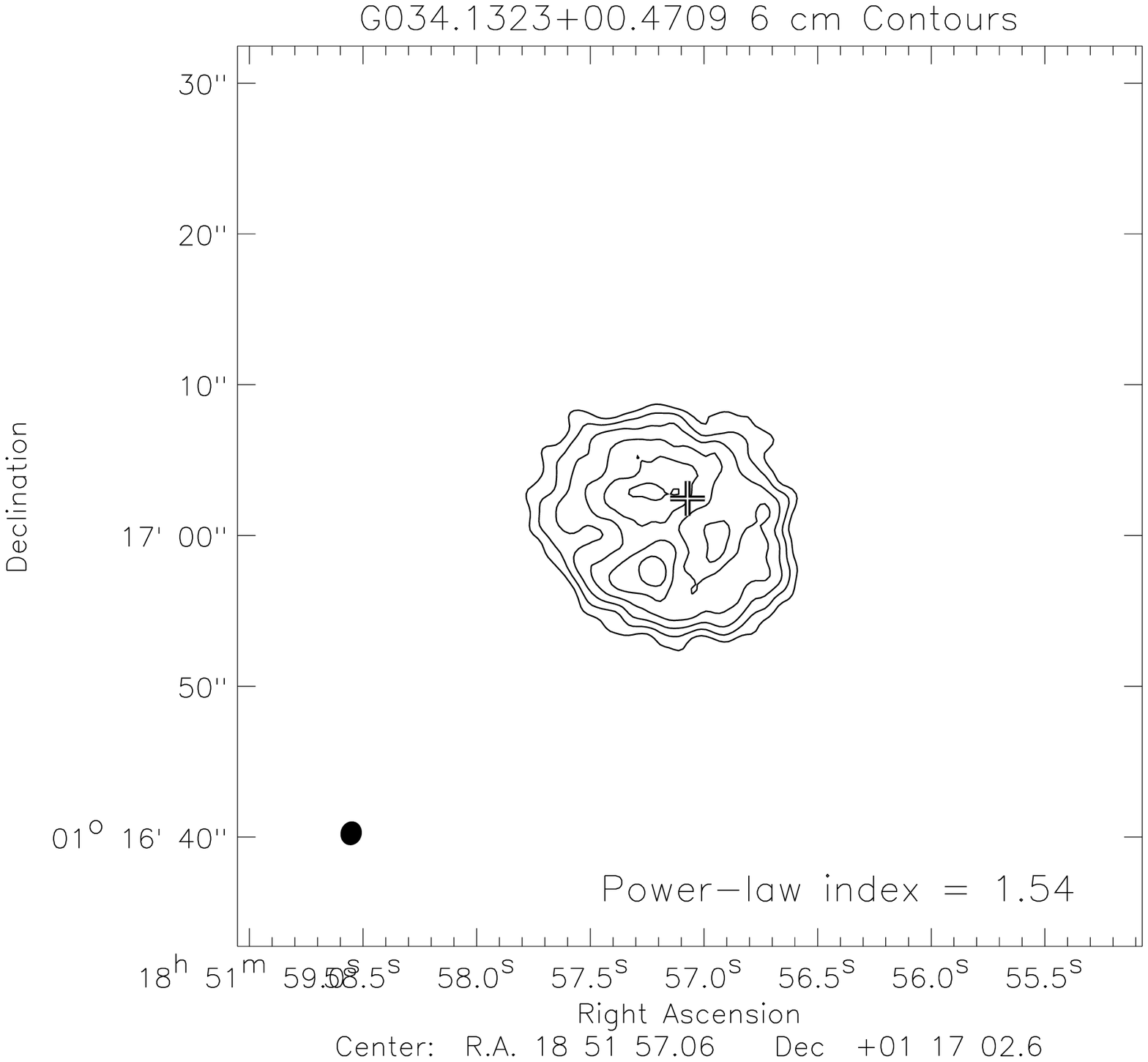}

\includegraphics[width=0.32\linewidth, trim=0 0 150 0]{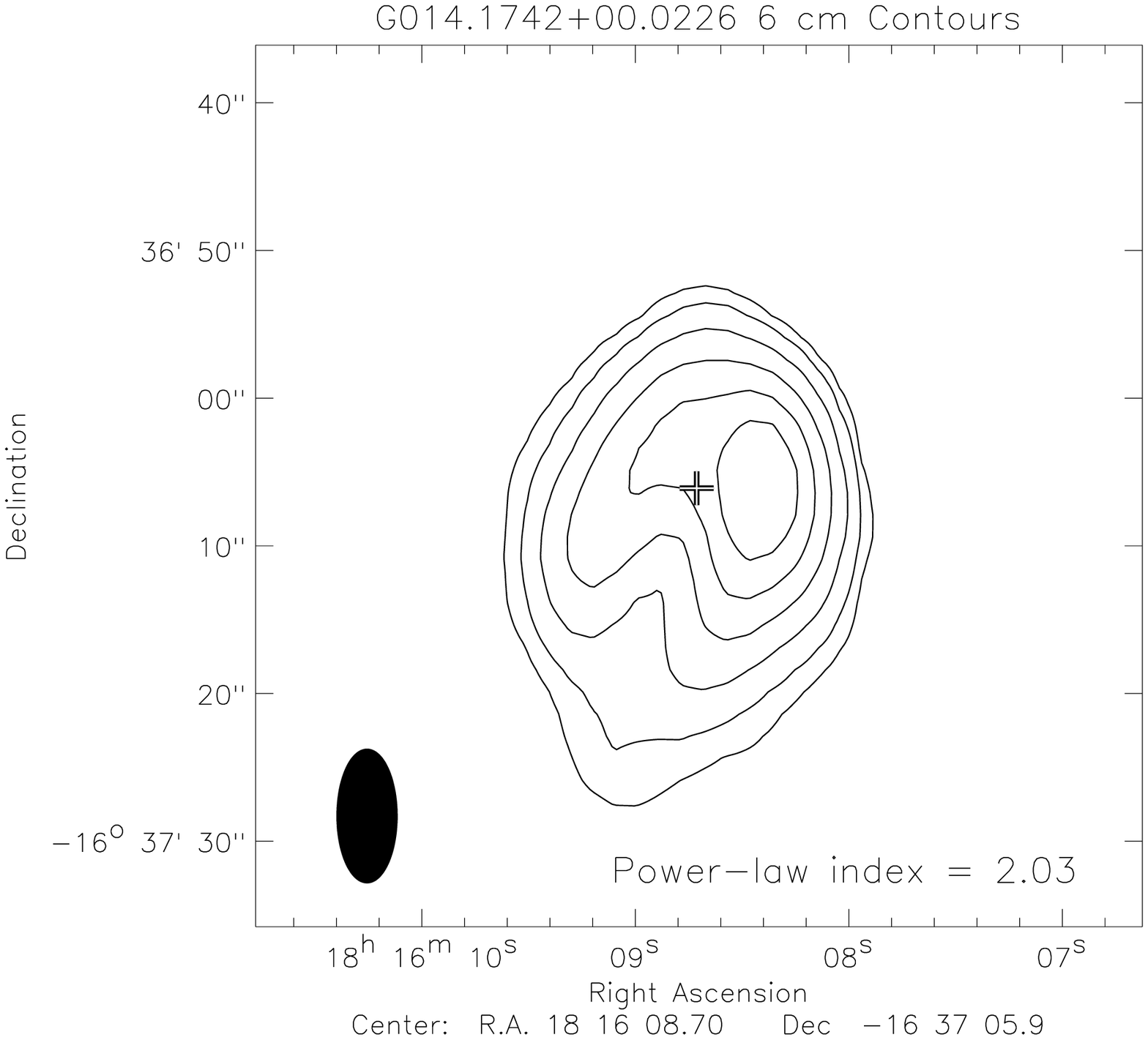}
\includegraphics[width=0.32\linewidth, trim=0 0 150 0]{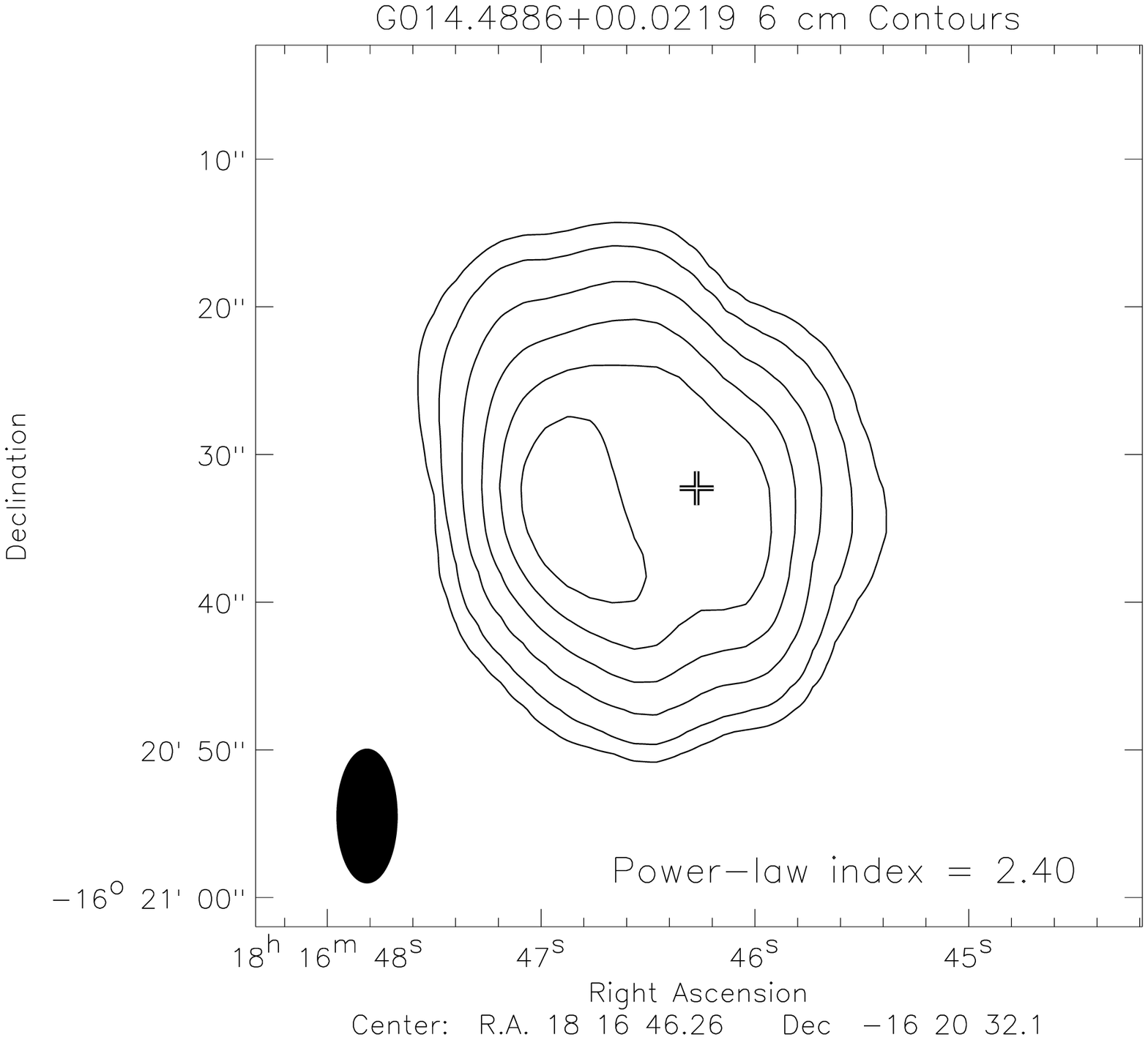}
\includegraphics[width=0.32\linewidth, trim=0 0 150 0]{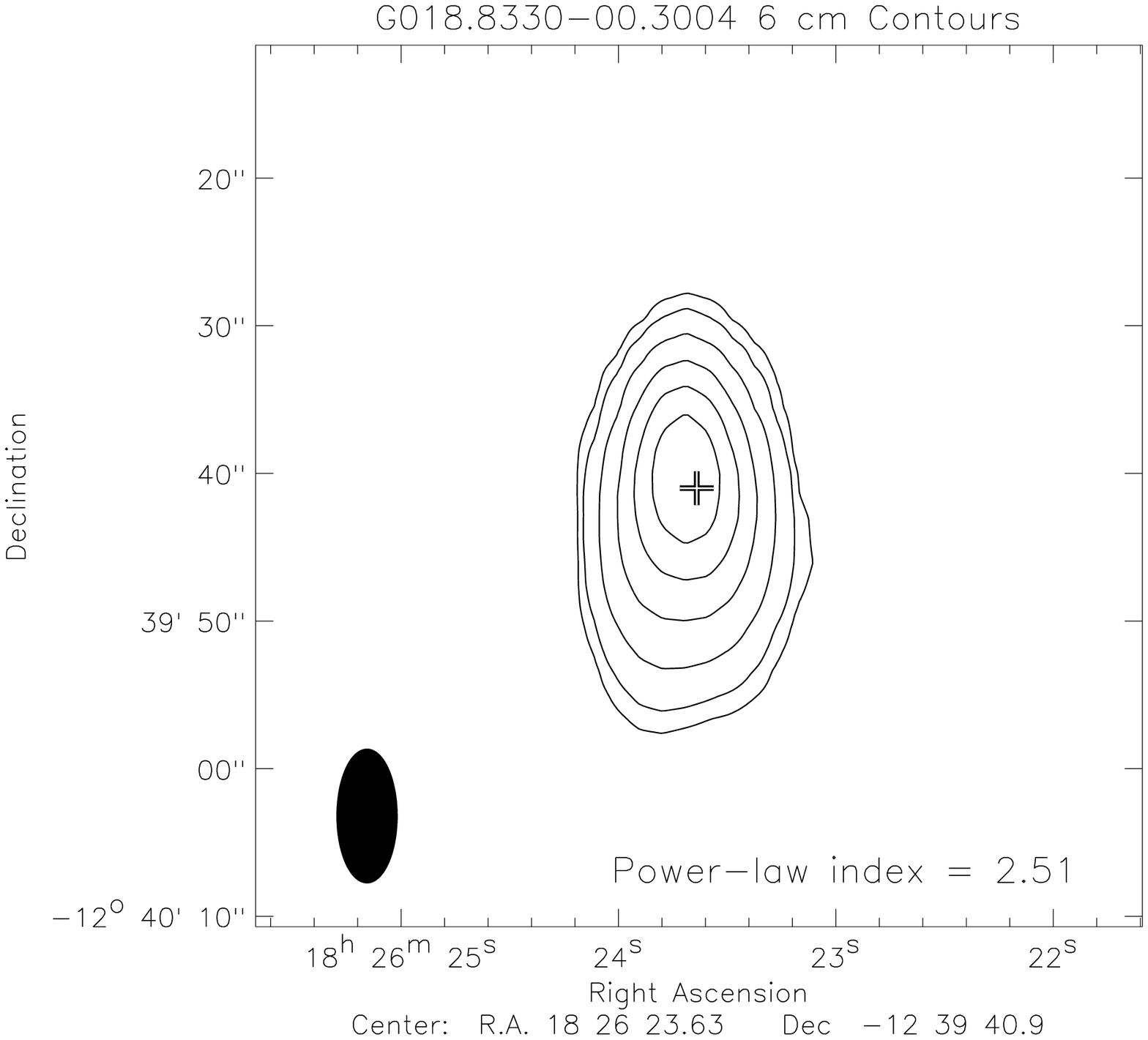}

\caption{\label{fig:rms_maps} Contour maps of the RMS-radio matches found from the targeted VLA observations (upper panels) and those found in the \citet{white2005} catalogue (lower panels). The first contour starts at 4$\sigma$ with the intervening levels determined by a dynamic power-law which is given in the lower right corner of each plot (see text for details). The RMS name is shown above the image and the size of the synthesised beam is shown to scale in the lower left hand corner. The position of the RMS source is marked with a cross. The full version of this figure is only available in electronic form at the CDS via anonymous ftp to cdsarc.u-strasbg.fr (130.79.125.5) or via http://cdsweb.u-strasbg.fr/cgi-bin/qcat?J/A+A/.}

\end{center}
\end{figure*}

In Fig.~\ref{fig:rms_maps} we present
contour maps of the distribution of radio emission found towards all
of RMS-radio matched sources. As with Paper I, the contour levels in each map have been determined using the dynamic range power-law fitting scheme described
by \cite{thompson2006}. The advantage of this scheme over a linear
scheme is in its ability to emphasise both emission from diffuse
extended structures with low surface brightness and emission from
bright compact sources. The contour levels were determined using the
following relationship $D=3\times N^i+4$, where $D$ is the dynamic
range of the map (defined as the peak brightness divided by the map's
r.m.s. noise), $N$ is the number of contours used (6 in this case), and
$i$ is the contour power-law index. Note this relationship has been
altered slightly from the one presented by \citet{thompson2006} so
that the first contours start at 4$\sigma$ rather than 3$\sigma$ used
by them. The lowest power-law index used was one, which resulted in
linearly spaced contours starting at 4$\sigma$ and increasing in steps
of 3$\sigma$.

The integrated flux values range from $\sim$1.3 mJy up to several Jy,
significantly above what would be expected from an ionising stellar
wind ($\sim$1~mJy at 1~kpc; \citealt{hoare2002}). These sources are
therefore unlikely to be genuine MYSOs and more likely to consist of
embedded compact and UCHII regions, and a small number of
PNe. However, to avoid the possibility that a small number of these
detections are actually due to ionised stellar winds in nearby sources
we will, once the high resolution far-infrared data and reliable distances becomes available, use the ratio of the radio and infrared luminosities (i.e., Log L$_{\rm{radio}}$/L$_{\rm{IR}}$; see Fig.~6 of  \citealt{hoare2007})  and
retrospectively apply the following criteria: UCHII region $>$ 8,
ionised stellar wind $<$ 8. Although this works well for sources with luminosities $>$ 10$^4$~\lsun, it is not so clear cut for weaker B stars where infrared spectroscopy will be needed to distinguish between stellar winds and UCHII regions.

Although this effectively eliminates a quarter of the RMS sources observed, we are still left with a large sample of MYSO candidates (702 sources or 72\% or the sample), and this is after eliminating two major sources of contamination. 

\subsection{Detection statistics}
\label{sect:detection_statistics}

\begin{figure}
\begin{center}
\includegraphics[width=0.45\textwidth]{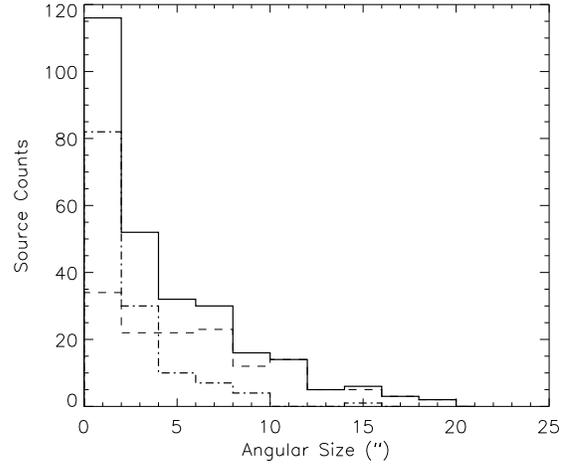}
\caption{\label{fig:angular_sizes} Histogram plots of the deconvolved angular sizes of the radio sources associated with RMS sources. The distribution of the whole sample is indicated by the histogram outlined by a solid line. The individual angular distributions of the targeted VLA and \citet{white2005} RMS-radio matches are outlined by a dashed-dotted and dashed line respectively. These histograms have been created using a 2\arcsec\ bin size.}

\end{center}
\end{figure}

The overall fraction of RMS sources with detected radio emission is similar to
the fraction found from the ATCA observations of southern sources
($\sim$25\%). However, breaking down the percentages between the two
catalogues we find a significantly higher fraction of matches are
found in the \citet{white2005} catalogue, i.e., the RMS-radio matches
from the targeted VLA observations $\sim$19.3\% compared to 46.5\% matched
using \citet{white2005} catalogue.

There are two reasons for the large difference in detection
rates. Firstly, the \citet{white2005}  survey region covers the busiest part of
the Galactic mid-plane where the majority of MYSOs and UCHII regions
are expected to be found -- the scale height of massive stars is
$\sim$0.5\degr\ (\citealt{reed2000}). Moving out of the mid-plane to
higher latitudes the sample is likely to contain a higher proportion
of evolved stars which have a much broader latitude
distribution. Secondly, the targeted VLA observations, due to the array
configuration and the limited \emph{uv} coverage, are in theory only
sensitive to angular scales of $\sim$20\arcsec, however, the \citet{becker1994}
6~cm survey, which mainly used the VLA C configuration, is sensitive
to scales as large as $\sim$150\arcsec. This point is nicely illustrated in the histogram plot of the angular sizes of sources presented in Fig.~\ref{fig:angular_sizes} and in the contoured emission maps presented in Fig.~4. Fig.~\ref{fig:angular_sizes} shows the geometrical deconvolved angular diameters of RMS-radio matches found from the targeted VLA observations and those found in the \citet{white2005} catalogue.

Although in theory the targeted VLA observations should have been sensitive enough to detect a source up to 20\arcsec\ in size, in practice these were only sensitive to sources roughly half this size due to the snapshot nature of the observations. There are a few points to bear in mind that arise from the limited sensitivity to angular scales larger than $\sim$10\arcsec. First, the targeted VLA observations have effectively filtered out emission from more extended
HII regions and so the number of HII regions identified should be
considered a lower limit. These more extended HII regions will be
identified through our programme of mid-infrared imaging (e.g.,
\citealt{mottram2007}). Second, care must be taken when comparing the
properties of the whole sample (e.g., Galactic distribution).

\subsection{Upper limits for non-detections}

\begin{figure}
\begin{center}
\includegraphics[width=0.45\textwidth]{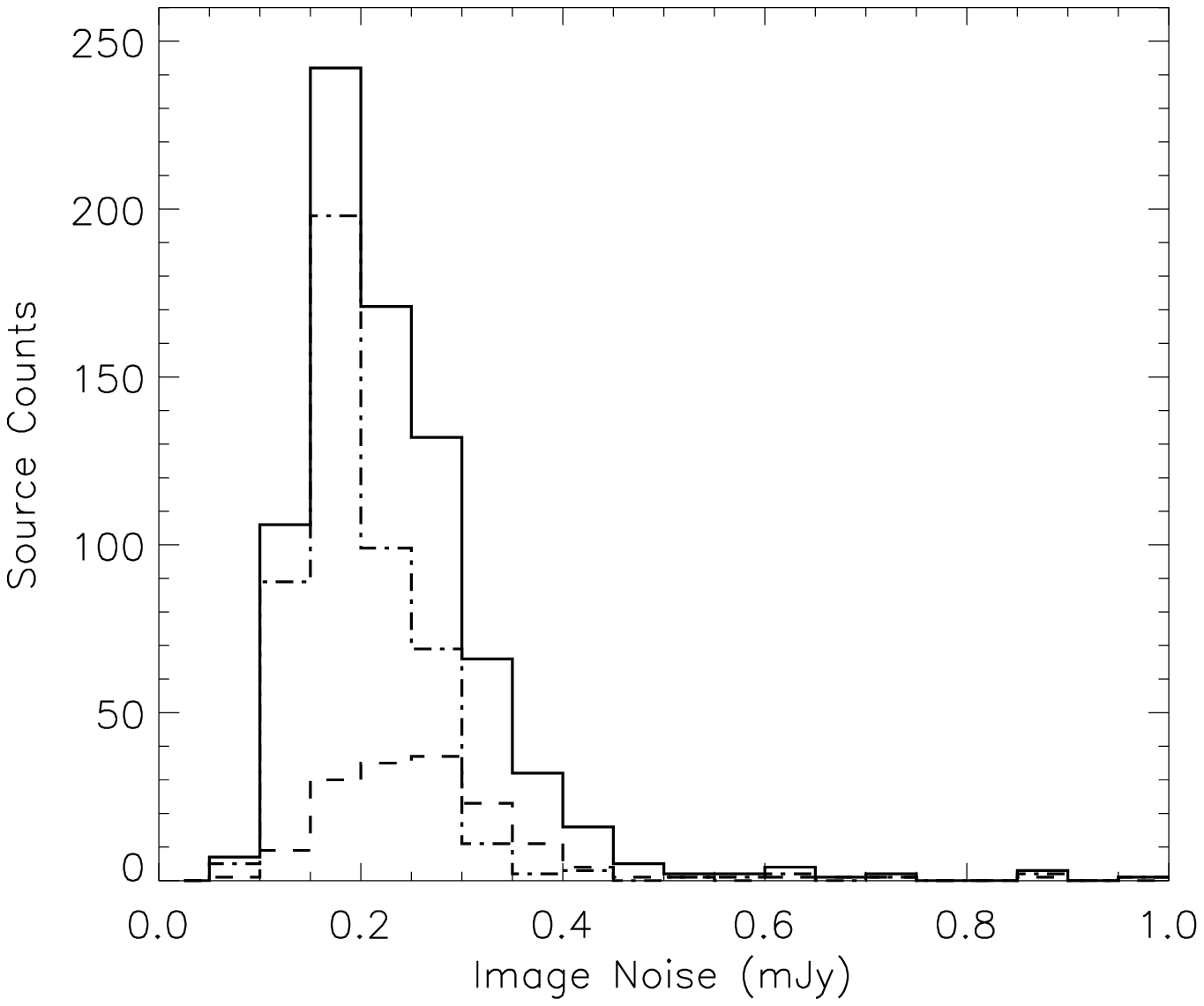}

\caption{\label{fig:local_rms} Histogram plots of the noise distribution
towards RMS radio non-detections. The distribution of the whole sample
is outlined by a solid line while the individual
noise distributions of the targeted VLA observations and the \citet{white2005} survey RMS radio non-detections matches are outlined by a dashed-dotted line and a dashed line respectively. These histograms have been created using a 0.05~mJy
beam$^{-1}$ bin size.}

\end{center}
\end{figure}

From the point of view of identifying a large sample of MYSOs it is the RMS
radio non-detections that are more interesting since they remain MYSO 
candidates. It is therefore useful to determine upper limits
for the radio flux towards these RMS sources. In
Fig.~\ref{fig:local_rms} we present a histogram of the measured local
noise levels at the position of the RMS sources where no radio
emission is detected (the noise distributions of the targeted VLA observations and the \citet{white2005} survey are also shown). The average 1$\sigma$ r.m.s. noise level is $\sim$0.22~mJy beam$^{-1}$ for both data sets. We have truncated the x-axis at 1~mJy beam$^{-1}$, however, the noise level only exceeds this for 47 sources ($\sim$6\%) of the radio non-detections. 

Using a 4$\sigma$ upper limit ($\sim$1~mJy) we find these observations
are sensitive enough to detect any optically thin ionised nebula
around a B0.5 or earlier type star across the Galaxy ($>$2 mJy at
$\sim$20~kpc; \citealt{kurtz1994}). We can therefore be reasonably
confident that the HII regions identified with these data presented here
is complete to all \emph{unresolved} optically thin HII regions driven by a high mass star with a luminosity of $>$10$^4$~\lsun\ located on the far side of the
Galaxy.  However, we are less sensitive, and therefore less complete, to optically thick, compact hyper-compact HII regions and more extended HII regions. However, the extended HII regions will be identified by their mid-infrared emission (e.g., \citealt{mottram2007}).

\section{Discussion}

\begin{figure*}
\begin{center}

\includegraphics[width=0.42\linewidth, trim=0 0 150 0]{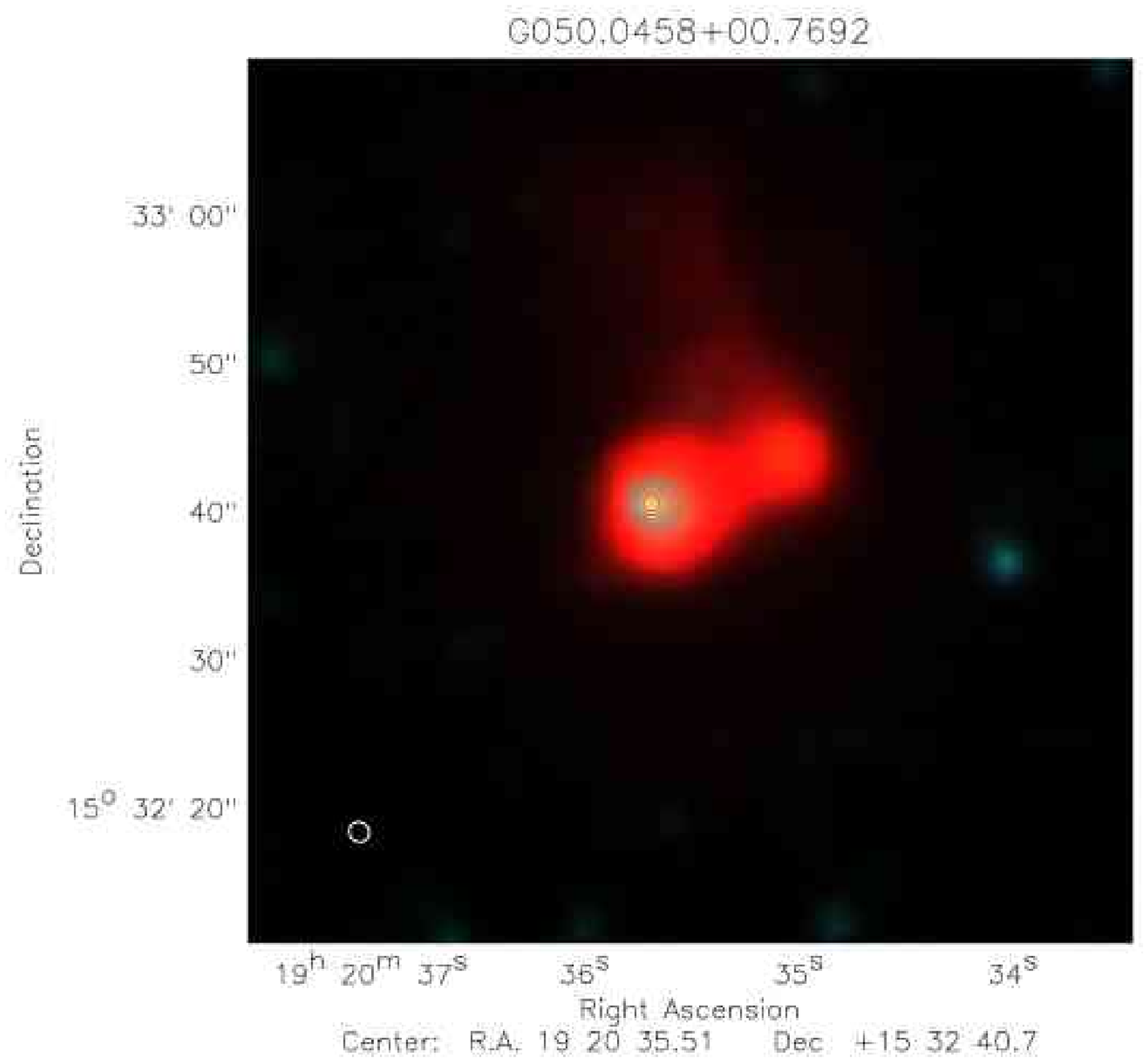}
\includegraphics[width=0.42\linewidth, trim=0 0 150 0]{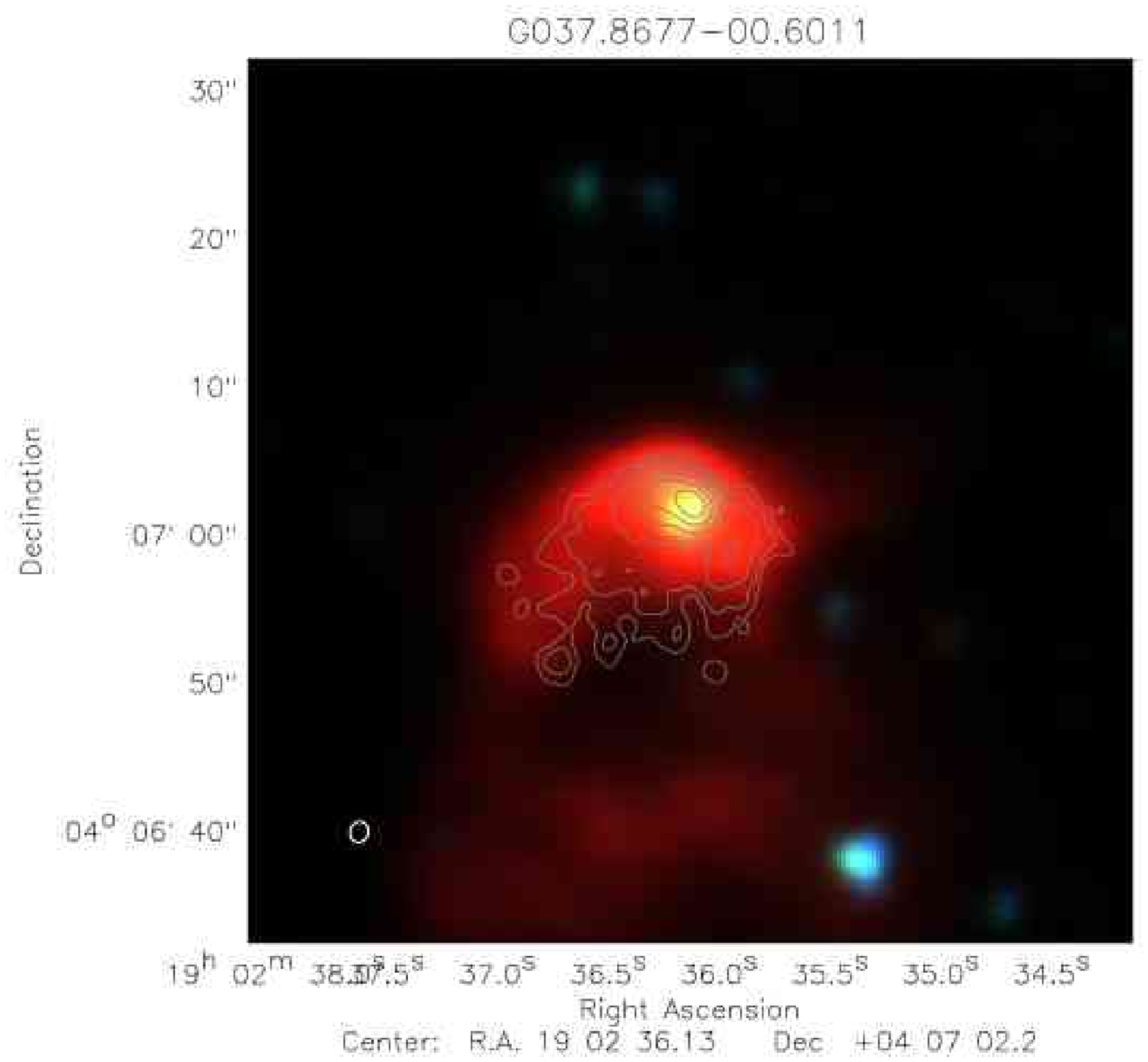}\\

\includegraphics[width=0.42\linewidth, trim=0 0 150 0]{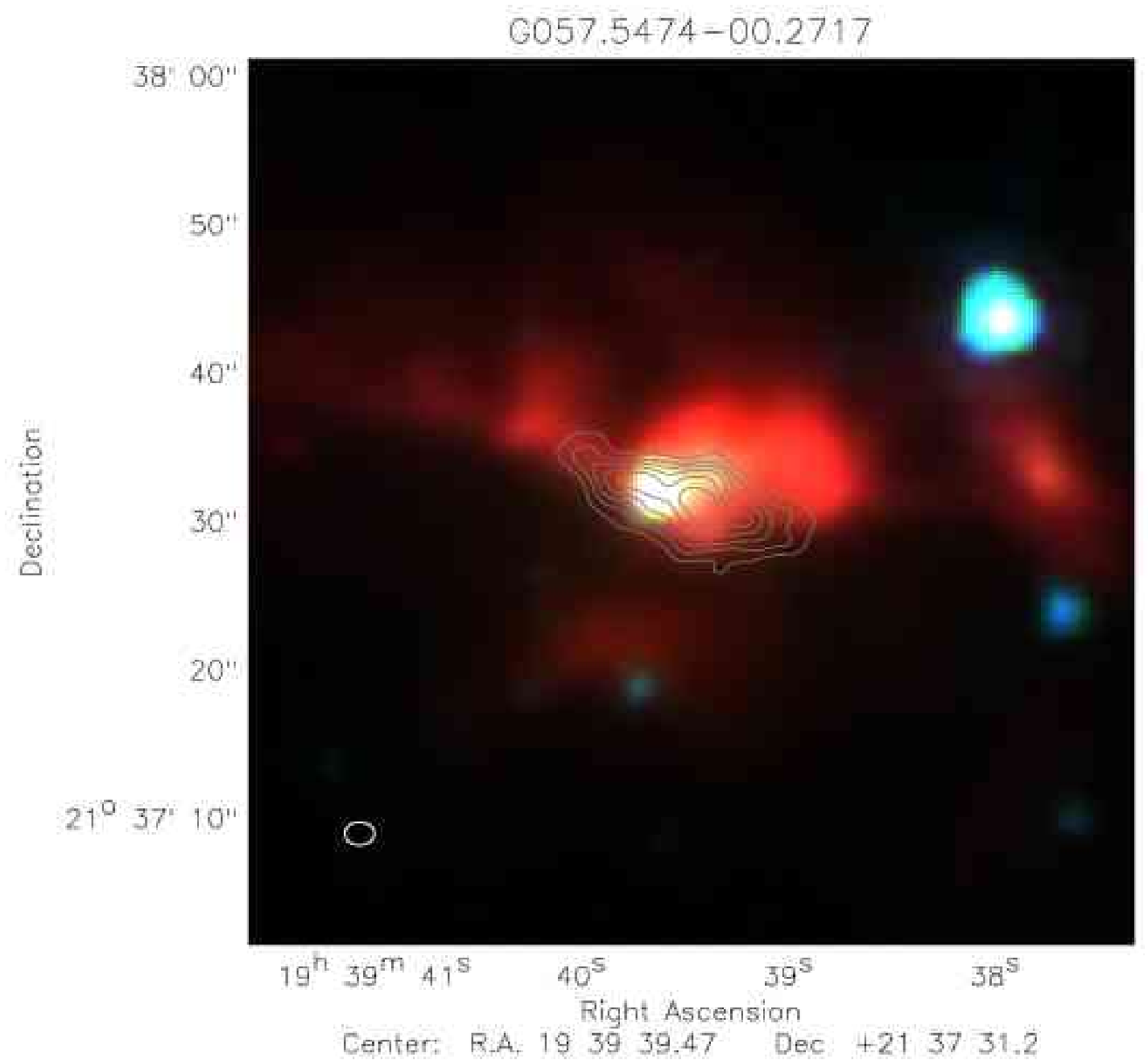}
\includegraphics[width=0.42\linewidth, trim=0 0 150 0]{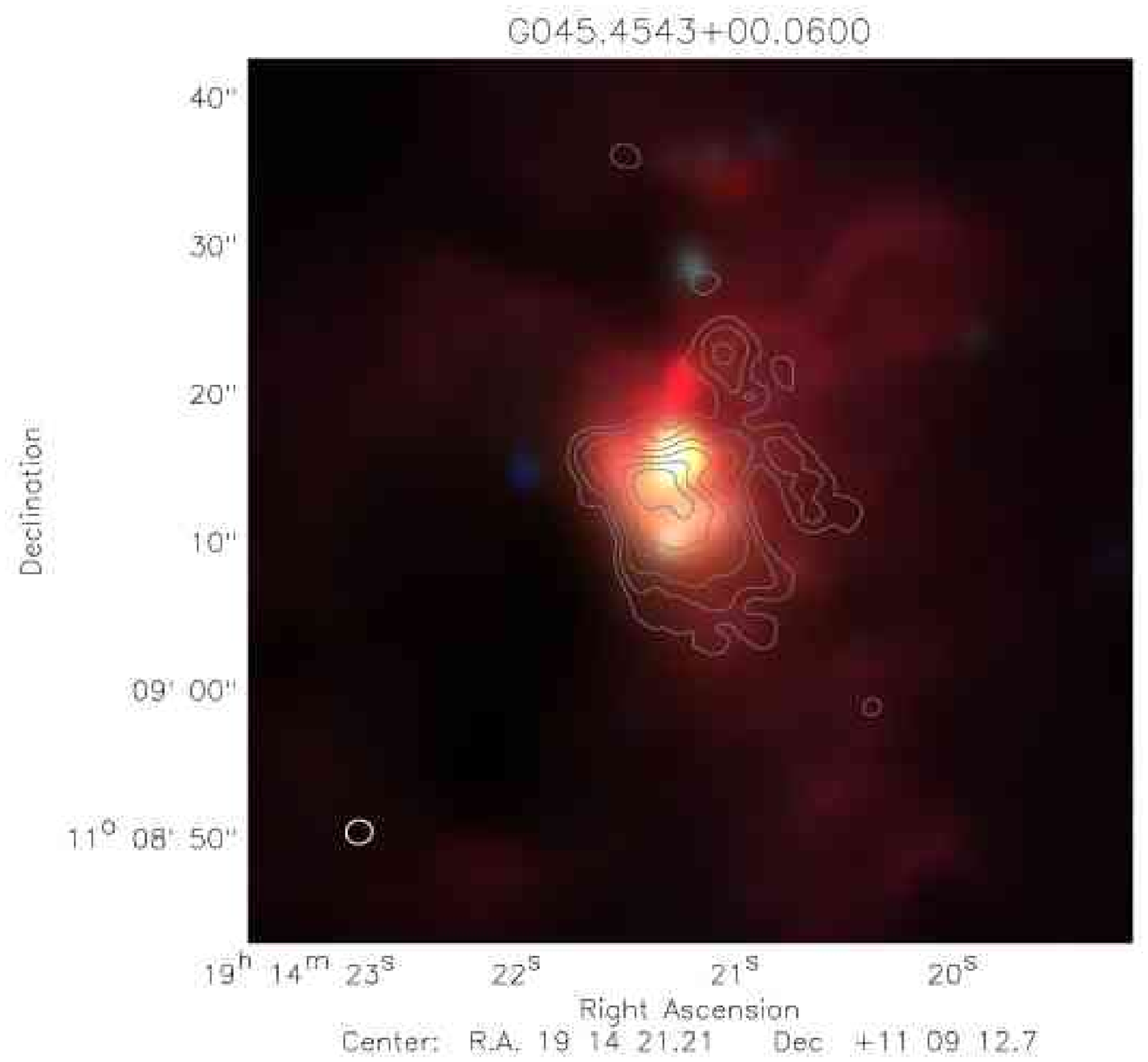}\\

\includegraphics[width=0.42\linewidth, trim=0 0 150 0]{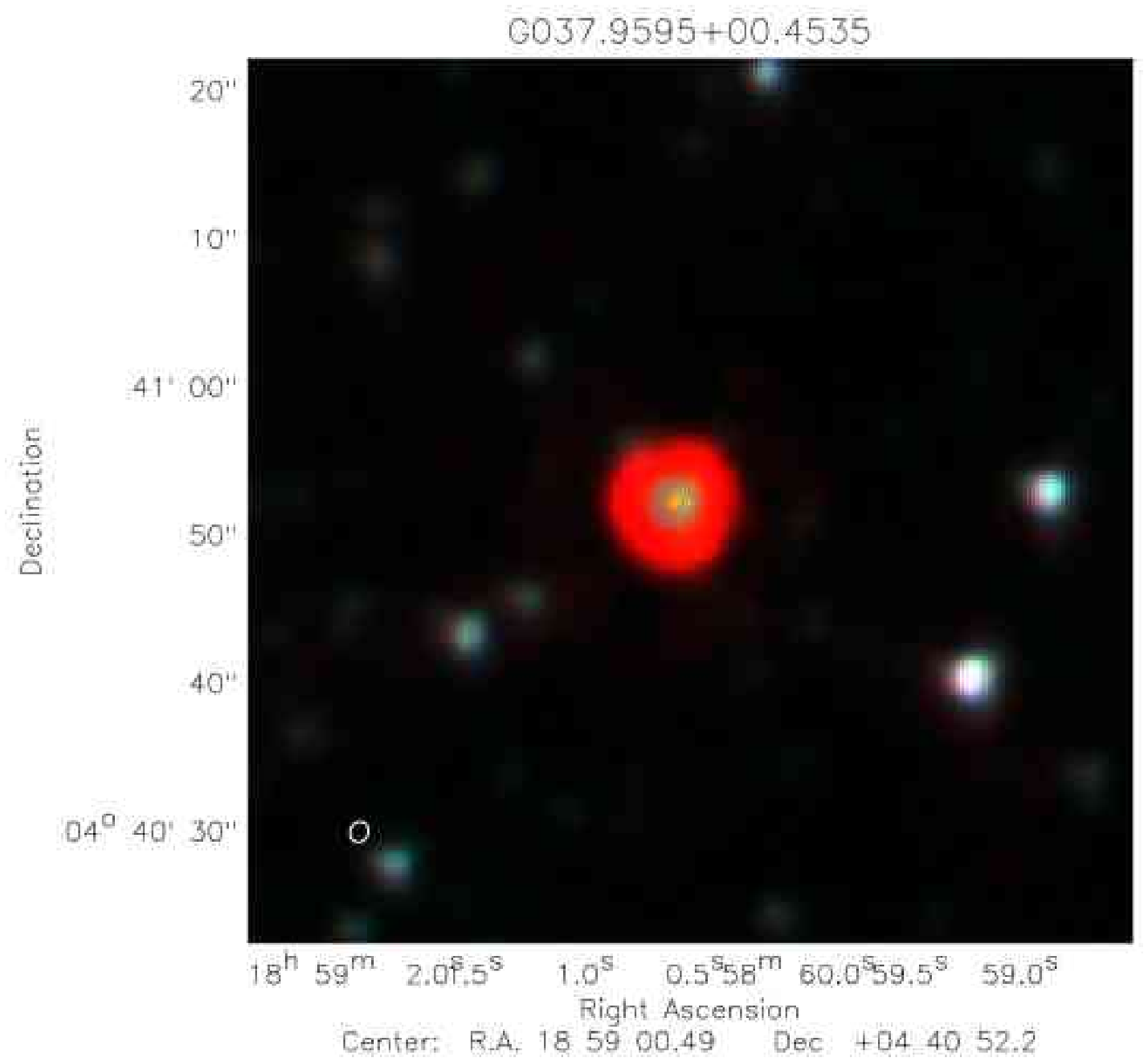}
\includegraphics[width=0.42\linewidth, trim=0 0 150 0]{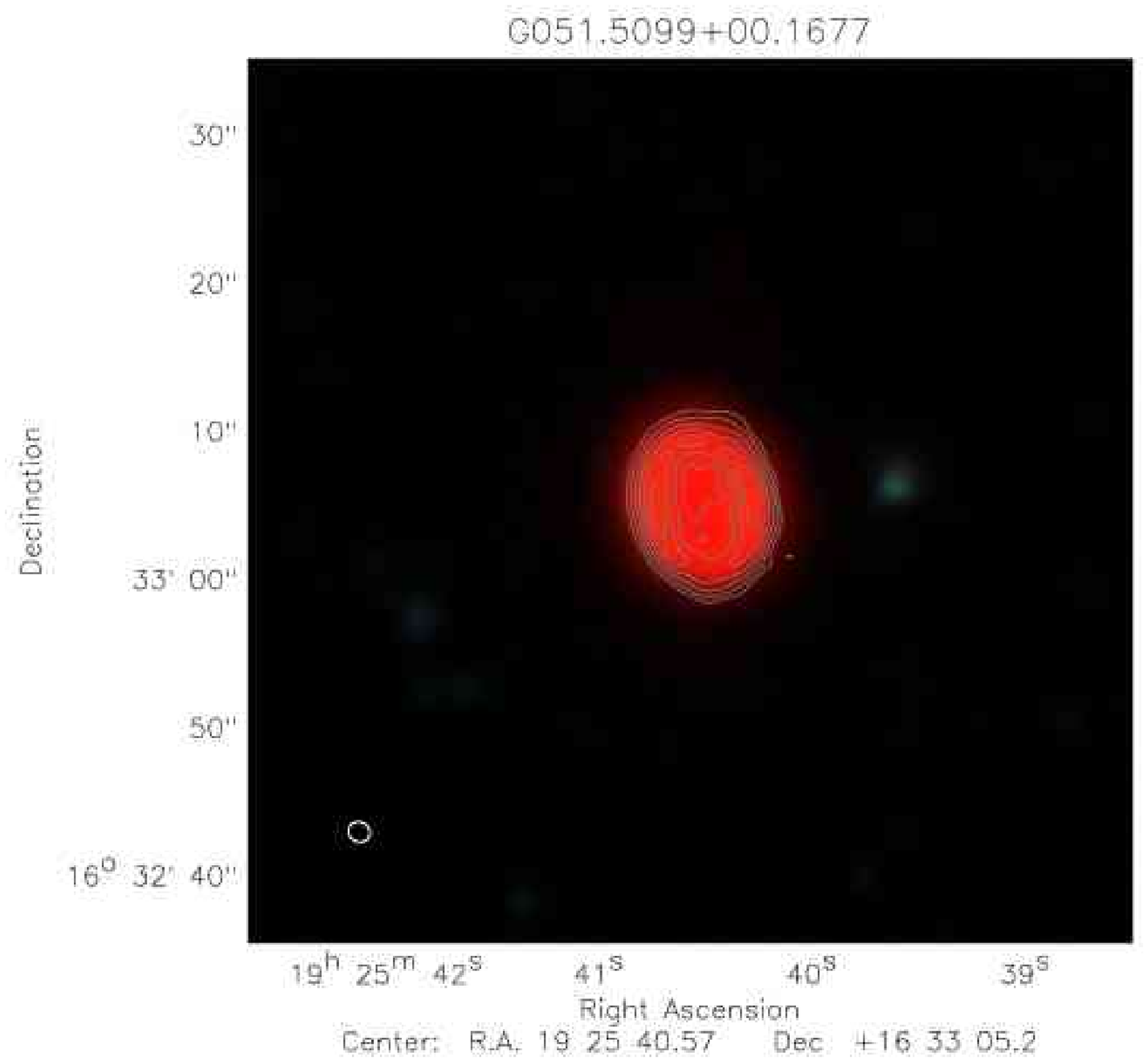}\\
	
\caption{\label{fig:glimpse_maps2} Examples of the different types of radio loud RMS sources identified. In the top and centre panels we present four HII regions chosen to illustrate the difference kinds of morphologies identified and in the bottom two panels we present an unresolved (left) and resolved (right) PN. These images are composed of the GLIMPSE 3.6, 4.5 and 8.0~\mum\ bands (coloured blue, green and red respectively) which are overlaid with contours of the detected radio eimission (contour levels are as for Fig.~\ref{fig:rms_maps}). The full version of this figure is only available in electronic form at the CDS via anonymous ftp to cdsarc.u-strasbg.fr (130.79.125.5) or via http://cdsweb.u-strasbg.fr/cgi-bin/qcat?J/A+A/.}

\end{center}
\end{figure*}

Kinematic distances are required before we can begin to analyses the
physical properties of these RMS-radio sources. We have recently
published the result of $^{13}$CO observations made towards both the
northern and southern hemisphere RMS sources
(\citealt{urquhart_13co_south,urquhart_13co_north}), however, we were
unable to unambiguously identify a velocity component to every
source. Moreover, for sources located within the solar circle the
Galactic rotation models give two possible distances equally spaced on
either side of the tangent point; this is known as the kinematic
distance ambiguity.  In this paper we will therefore restrict our discussion to the investigation of the multiple radio detections, the identification of PNe and compact HII regions and their Galactic distribution. 

The resolution of the MSX data is too low to effectively investigate the relationship between the mid-infrared emission from warm dust and radio emission from ionised gas. However, a large proportion of these sources are located within a region surveyed by Spitzer as part of the GLIMPSE legacy project
(\citealt{benjamin2003}). Therefore, for a number of these sources high
resolution mid-infrared imaging data are available that allow us to
study them in greater detail. We have combined the 3.6, 4.5 and 8.0 $\mu$m band images obtained with the IRAC camera (\citealt{fazio2004}) to create three colour images (coloured blue, green and red respectively) for all RMS-radio matches within the region of the Northern Galactic Plane imaged by GLIMPSE
($10\degr< l <65\degr$ and $|b| < 1$). We have over-plotted contours
of the observed radio emission on these images in order to compare the
distribution of the infrared and radio emission. 

In Fig.~\ref{fig:glimpse_maps2} we present six examples of the two kinds of RMS-radio sources found in the sample and the different types of morphologies associated with them. We have selected these examples from the targeted VLA observations as their higher resolution is more closely matched to that of GLIMPSE data allowing the spatial features to be compared at similar resolutions.   In the top and centre panels we present examples of HII regions chosen to illustrate the various types of morphologies identified (clockwise from the top left): unresolved, cometary, multipeaked and irregular. In the two bottom panels of this figure we present images of an unresolved (left) and resolved (right) PNe.

\subsection{Distinguishing between HII regions and PNe}
\label{sect:classification}

So far we have discussed RMS sources that are associated with radio
emission, however, not all of these are HII regions. A small number of
PNe still remain in the sample and these need to be identified and
removed before we can properly investigate the distribution and
statistical properties of the HII regions. The radio and MSX data
alone are not sufficient to separate these two kinds of objects.
However, by combining data from other parts of our multi-wavelength
programme of observations it is possible to distinguish between them.

PNe have similar mid-infrared colours to UCHII regions and
YSOs, and in the case of UCHII regions are both associated with radio
emission. However, they, like other types of evolved stars, are not associated
with large amounts of molecular gas. Hence, evolved stars and PNe are not generally associated with strong CO emission ($^{12}$CO $J$=1--0 and $J$=2--1 typically less than 1~K; \citealt{loup1993}). As part of our campaign of follow-up observations we have conducted a programme of $^{13}$CO observations
towards all RMS sources (\citealt{urquhart_13co_south,urquhart_13co_north}). The main focus of these molecular line observations is to derive kinematic distances to the sources, which can then be used to estimate luminosities and
determine galactic location. However, the non-detections also allow us to 
identify and eliminate a significant number of evolved stars.

Evolved stars tend to appear isolated when imaged at near and
mid-infrared wavelengths and look almost identical in all wavebands
(JHK and 3.6, 4.5, 5.8 and 8.0~\mum), whereas YSOs and UCHIIs are
often associated with nebulous emission and regions of extinction,
and/or found in small clusters, which leads to observable differences
in the infrared images. The final piece of information we can use to help identify evolved stars is their far-infrared spectral energy distribution (SED). Unlike the SEDs of YSOs and HII regions that both peak at approximately
100~\mum, the SEDs of evolved stars tend to peak at shorter
wavelengths. A number of evolved sources can therefore be identified
from by their IRAS fluxes, which are seen to peak at either 25 or
60~\mum\ (\citealt{manchado1989}).

Using a combination of data collected as part of the multi-wavelength
programme and archival infrared images we have
classified 51 of the 272 RMS-radio matches as PNe. This makes up just
$\sim$20\% of the RMS-radio sample. Having removed all of the PNe we
are left with a clean sample of 208 genuine compact and ultra compact
HII regions. To illustrate the differences between PNe and HII regions we present examples of each in the Fig.~\ref{fig:glimpse_maps2}. These sources have been chosen as they exemplify some of the differences between the two classes of objects discuss in the previous paragraphs. The PNe are isolated and are not associated with extensive diffuse and/or irregular nebulosity, or extinction lanes, which would be expected if these were associated with molecular material. In stark contrast the HII regions are associated with large, extended regions of nebulosity, and appear to be surrounded by regions of extinction.

In the previous paragraphs we have discussed how we have distinguished between PNe and HII regions and have successfully identified 259 RMS sources as either one type or the other. However, this leaves thirteen sources so far unaccounted for. Seven of these sources have been identified as YSOs; for these sources we are currently unable to determine if the radio emission indicates the presence of a stellar wind, a radio jet or is the result of a chance alignment.  Additional high resolution radio observations will be required to investigate these possibility further. The remaining six sources are an adhoc mixture of various types of radio stars and galaxies previously identified in the literature, or from inspection of the multi-wavelength data stored in the RMS database; these are classified as `Other' in Table~5.\footnote{More detailed explanations concerning source classification can be found on the RMS database.}

\subsection{Distribution of RMS-radio associations}

In this subsection we examine the Galactic distributions of the PNe and HII regions. In the previous section we identified a large number of PNe and HII regions, however, to avoid any selection effects, particularly due to the filtering out of larger HII regions 
(see Sect:~\ref{sect:detection_statistics}), we now include all HII
regions and PNe identified either from the literature or from our multi-wavelength data. The total number of northern HII regions and PNe so far identified are 391 and 79, respectively. Our ongoing classification of sources is not yet complete as there is not yet sufficient data available to make
a determination for a number of sources ($\sim$10\%) and therefore
these numbers may change slightly. Including the HII regions and PNe detected from the ATCA observations (Paper~I) we might expect to identify $\sim$600--800 HII regions and $\sim$170-200 PNe in the RMS sample as a whole.

\begin{figure*}
\begin{center}
\includegraphics[width=0.95\textwidth, trim= 50 0 50 0]{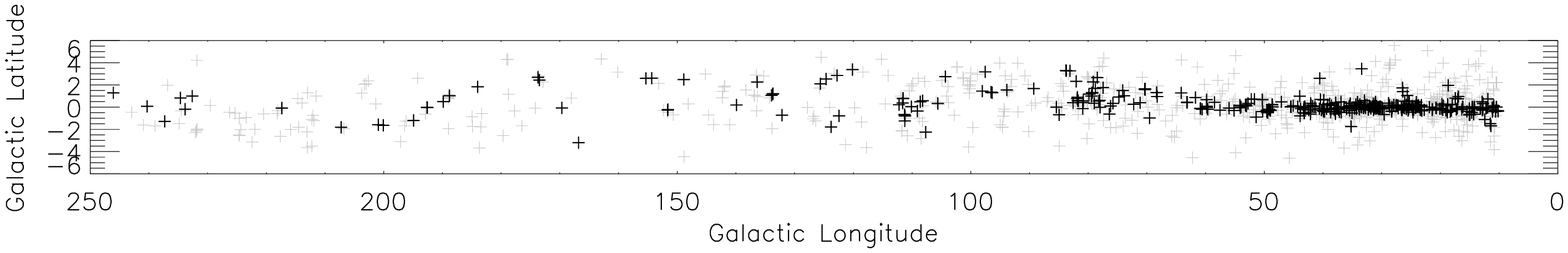}
\includegraphics[width=0.95\textwidth, trim= 50 0 50 0]{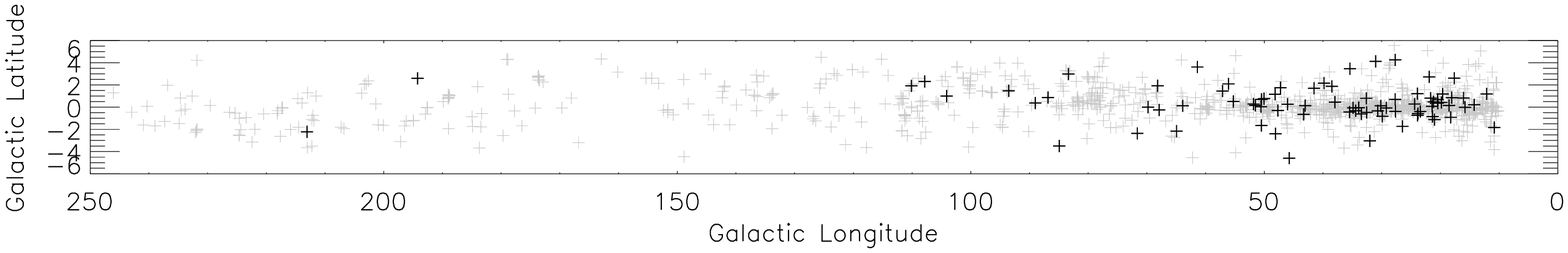}
\includegraphics[width=0.95\textwidth, trim= 50 0 50 0]{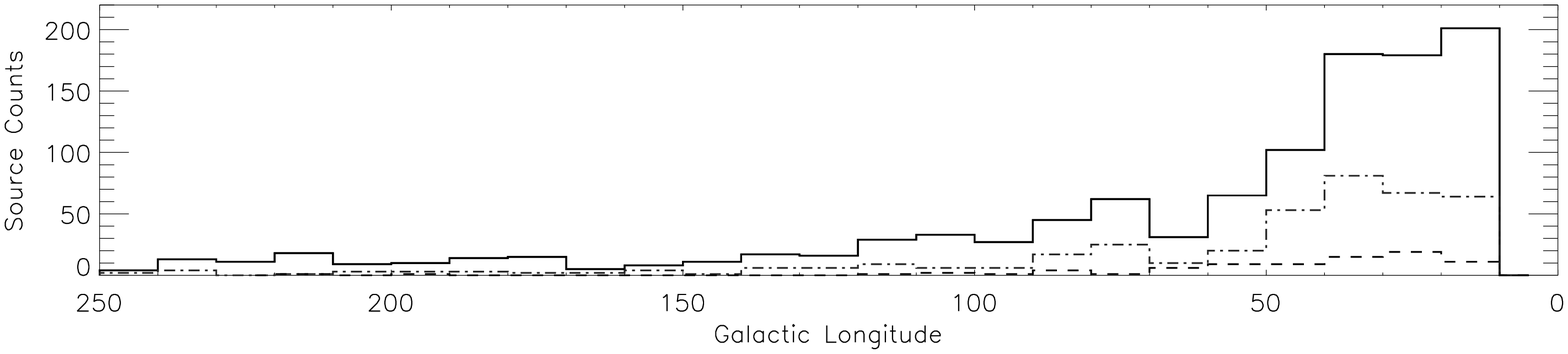}

\caption{\label{fig:uchii_long_lat_distribution} Galactic distribution
of all RMS sources observed as part of the targeted VLA observations or located within the region surveyed  by \citet{white2005}. In the top two panels we mark the positions of all RMS sources observed by grey crosses as a function of longitude and latitude. We indicate the positions of all HII regions and PNe by overplotting black crosses in the top and middle panels respectively. In the lower panel we present a histogram plot of the number of RMS source observed (solid line), and HII regions (dashed-dotteds line) and PNe (dashed line) detected as a function of Galactic longitude.  }

\end{center}
\end{figure*}

In Fig.~\ref{fig:uchii_long_lat_distribution} we present plots of the Galactic longitude-latitude distribution of all northern hemisphere RMS sources. In the top two panels we mark the positions of all RMS sources observed and the positions of all HII regions and PNe identified. The distribution of HII regions is strongly correlated with the Galactic mid-plane, particularly at longitudes less than 60\degr\ where all but a handful of sources can be found within 1\degr\ of the plane. This region also contains the majority of the HII regions ($\sim$70\%). For longitudes larger than 50\degr\ the latitude correlation with the mid-plane becomes progressively worse and rapidly disappears. The
distribution of PNe is quite different to that of the HII
regions. Like the distribution of HII regions, they are also
concentrated towards the inner Galaxy, however, their latitude
distribution is fairly flat between $|b|$ $<$ 1 before quickly tailing
off above this value.

\begin{figure}
\begin{center}
\includegraphics[width=0.49\textwidth]{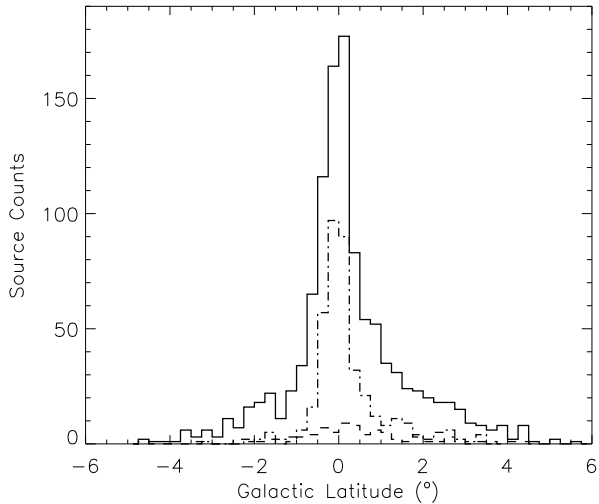}

\caption{\label{fig:uchii_lat_distribution} Histogram plots of the latitude distributions of all northern hemisphere RMS sources, all identified HII regions and PNe are outlined by solid, dashed-dotted and dashed lines, respectively. The data has been binned in units 0.25\degr.}

\end{center}
\end{figure}

The degree of correlation of HII regions and PNe with the Galactic mid-plane is
better illustrated in Fig.~\ref{fig:uchii_lat_distribution}. In this
figure we present a histogram of the number of RMS sources observed as
a function of Galactic latitude. In this figure we have also over-plotted the latitude distributions of the HII regions and PNe. The sample of HII regions is relatively clean of contamination, covers a broad Galactic latitude range and is more complete than previously reported samples. This sample is therefore probably the best representation to date of the Galactic population of HII regions as a whole.

 Using this sample we calculate the Galactic scale height to be
$\sim$0.6\degr. This is in line with the values derived from the UCHII regions identified from the ATCA observations (Paper~I) and by \citet{reed2000} from the
distribution of local OB stars ($\sim$0.5\degr). The Gaussian FWHM of the latitude distribution is $\sim$0.7\degr, which is almost three times larger than that reported by \citet{white2005} from their 6~cm survey ($\sim$0.25\degr), however, their survey only extended out to $|b| < 0.4$\degr\ and so they have missed a significant number of HII regions which has led to an underestimation of the true FWHM (see also discussion by \citealt{kurtz1994}).

\section{Summary and conclusions}
\label{sect:summary}

These observations form part of a multi-wavelength programme of
follow-up observations of a sample of $\sim$2000 colour selected MYSO
candidates designed to distinguish between genuine MYSOs and other
embedded or dusty objects. In this paper we report the results of
radio continuum observations towards a sample of MYSO candidates
located in the northern sky (10\degr $<$ $l$ $<$
250\degr). Observations were made using the VLA at a frequency of
5~GHz ($\sim$6~cm) towards 659 RMS sources. In addition to the fields
observed we present archival radio data for a further 315 RMS sources obtained
as part of the \citet{white2005} 6~cm survey of the inner Galactic plane.

Combining these two radio data sets provide continuum data for 974 RMS
sources ($\sim$50\% of the sample).  These observations were aimed at
identifying radio loud contaminants such as UCHII regions and PNe from
the relatively radio quiet MYSOs. These observations have a typical
rms noise value of $\sim$0.2~mJy and are therefore sensitive enough
to have detected an unresolved HII region powered by B0.5 or earlier type star located at the far side of the Galaxy. Our main finding are as follows:

\begin{enumerate}

\item The targeted VLA observations detected 669 radio sources with fluxes
above a nominal 4$\sigma$ cutoff ($\sim$1~mJy). Cross matching these
sources with the MSX point source catalogue we find mid-infrared
counterparts for approximately one third of the radio detection -
these are likely to be Galactic HII regions or PNe. The remaining
two-thirds of the radio detections for which no mid-infrared source
could be identified are thought to consist mainly of extragalactic
background objects.

\item Cross matching the VLA detections presented here, and the \citet{white2005} 6~cm catalogue, with the sample of MYSO candidates we identified 272 RMS sources that are associated with radio emission. In total we have
found that approximately $\sim$27\% of the sample is contaminated by
radio loud sources -- this compares well with the results of the
Australia Telescope Compact Array survey of the southern sample of
MYSOs which reported a $\sim$25\% radio association rate.

\item More interestingly from the point of view of detecting MYSOs we
failed to detect any radio emission towards 702 RMS sources, and
therefore after eliminating two of the main sources of contamination
we are still left with a large sample of MYSO candidates. Moreover,
we have measured the noise in these images and been able to place
4$\sigma$ upper limits of $\sim$1~mJy on these non-detections.

\item Using data obtained from other parts of our multi-wavelength
programme of follow-up observations and archival near-, mid- and
far-infrared data we have separated the RMS-radio matches into two
distinct types of objects: we have identified 208 HII regions and 51
PNe from the 272 RMS-radio matches. These numbers increase to 391 for
the HII regions and 79 PNe once we include back in these types of objects
previously identified in the literature which were excluded from these
observations.

\item We have identified 51 PNe which is $\sim$20\% of the radio
matches, however, taking into account the PNe identified in the
literature (28 PNe) and excluded from these observations we estimate
the total fraction of PNe that contaminate the RMS sample is approximately
10\%.

\item Using our unbiased sample of 391 compact and UCHII regions we
estimate the Galactic scale height of massive stars to be 0.6\degr. This is in line with the
values derived from the UCHII regions identified from the ATCA
observations (\citealt{urquhart_radio_south}) and by \citet{reed2000}
from the distribution of nearby OB stars.

\end{enumerate}

The radio continuum data presented here complement the observations of
the southern hemisphere sample of RMS sources presented in
\citet{urquhart_radio_south}. The full data set now includes 6~cm observations for
$\sim$1800 MYSO candidates, and together with archival data provides
radio continuum data for the entire RMS catalogue. In turn this data
set forms part of a larger programme of follow-up observations of near
and mid-infrared colour selected sample of MYSO candidates and are
significant step in the delivery of the largest sample of MYSOs to
date.

\begin{acknowledgements}

The authors would like to thank the Director and staff of the VLA for
their assistance during the preparation of these observations. The authors would also like to thank the referee Michael Burton for some useful comments and suggestions. JSU is partially 
supported by a STFC postdoctoral grant and a CSIRO fellowship. This research would not have been possible without the SIMBAD astronomical database service
operated at CDS, Strasbourg, France and the NASA Astrophysics Data
System Bibliographic Services. This research makes use of data
products from the MSX and 2MASS Surveys, which is are joint projects of the
University of Massachusetts and the Infrared Processing and Analysis
Center/California Institute of Technology, funded by the National
Aeronautics and Space Administration and the National Science
Foundation.

\end{acknowledgements}

\bibliography{2108}

\bibliographystyle{aa}




\end{document}